\documentclass[aps,pre,twocolumn,showkeys]{revtex4}
\usepackage{amsmath}
\usepackage{amsfonts}
\usepackage{amssymb}
\usepackage{graphicx}
\usepackage{dcolumn}
\usepackage{bm}

\newcommand{\w}{\omega}
\newcommand{\wb}{\omega_\mathrm{b}}
\newcommand{\Tb}{T_\mathrm{b}}
\newcommand{\we}{\omega_\mathrm{e}}
\newcommand{\F}{\mathcal{F}}

\begin{document}

\title{Polarobreathers in soft potentials}

\author{J. Cuevas$^1$, P.G.\ Kevrekidis$^{2}$, D.J. Frantzeskakis$^3$
and A.R. Bishop$^4$}
\affiliation{$^{1}$ Grupo de F{\'i}sica No Lineal, Departamento de
F{\'i}sica Aplicada I, Escuela Universitaria Polit{\'{e}}cnica, C/
Virgen de {\'{A}}frica, 7, 41011
Sevilla, Spain\\
$^{2}$ Department of Mathematics and Statistics, University of
Massachusetts, Amherst MA 01003-4515, USA \\
$^3$ Department of Physics, University of Athens, Panepistimiopolis,
Zografos, Athens 15784, Greece \\
$^4$ Theoretical Division and Center for Nonlinear Studies, Los Alamos
National Laboratory, Los Alamos, New Mexico 87545, USA}

\date{\today}

\begin{abstract}
We consider polarons in models of coupled electronic and vibrational
degrees of freedom, in the presence of a soft nonlinear
inter-particle potential (Morse potential). In particular, we focus
on a a bound state of a polaron with a breather, a so-called
``polarobreather''. We analyze the existence of this branch based on
frequency resonance conditions and illustrate its stability using
Floquet spectrum techniques. Multi-site solutions of this type are
also obtained both in the stationary case (two-site polarons) and in
the breathing case (two-site polarobreathers). We also obtain a
different branch of solutions, namely a polaronic nanopteron.
\end{abstract}



\maketitle

\section{Introduction}

In the past few years, the study of discrete nonlinear systems has
received a lot of attention, chiefly due to the increasing number of
pertinent physical applications \cite{reviews}. Among the various
areas of recent interest, one can highlight coupled waveguide arrays
in nonlinear optics \cite{reviews1}, Bose-Einstein condensates
(BECs) trapped in deep optical lattices (OLs) \cite{reviews2} in
atomic physics, coupled cantilever systems in nano-mechanics
\cite{sievers}, the local denaturation of the DNA double strand in
biophysics \cite{reviews3}, stellar dynamics in astrophysics
\cite{voglis}, and so on.

In the study of such discrete nonlinear systems, an important concept is the
possibility of intrinsic localization due to interacting
degrees of freedom, even in a linear regime, as proposed by Holstein
\cite{f9,f10}. This type of systems has been shown to sustain
single- and multi-humped polaronic and excitonic solutions \cite{f11}.
The same notion of effective nonlinearity in the
setting of coupled excitonic and vibrational degrees of freedom
was central to Davydov's suggestion of solitonic excitations
arising in biomolecules \cite{f3,f4}. More recently, a new aspect
has been added to this type of problems \cite{FUENTES,MKRB03},
by considering the interplay of the linear self-trapping with
a soft nonlinear potential, such as a Morse potential. This has been
proposed as a more general model relevant to many soft matter applications.

Our purpose in the present paper is to examine this type of model
and to provide a number of insights regarding its solutions and
their stability. We start by establishing that the main stationary
solutions obtained in the earlier work of \cite{FUENTES} (namely,
the single-site polarons) are dynamically stable within their region
of existence. We then turn our attention to a different class of
solutions which are genuinely ``breathing'' in time and which will,
thus, be termed {\em polarobreathers}. Such solutions were first
obtained and discussed in the context of the Holstein model in
\cite{KA98,A97}, from where this term is coined. Here, we will
obtain (and justify, based on resonance conditions) the domain of
existence of such solutions. We will also systematically investigate
their stability, by performing the corresponding Floquet spectral
analysis, and illustrate the existence of intervals of linear
stability of these solutions. Our study of this new class of
solutions will reveal yet another family of solutions consisting of
a static polaron combined with a linear mode to give a form of
polaronic nanopteron. Since the branch of stationary polarons
terminates at a critical value of the coupling as illustrated in the
earlier work of \cite{FUENTES}, we will examine the dynamics beyond
this termination point with a single hump initial condition and will
show that the solution develops large breathing fluctuations in its
local (i.e., central site) energy in time due to an interplay
between the central site of the localized profile and its neighbors.
We will also highlight multi-site branches such as the two-site
stationary solutions (polarons) and the two-site polarobreathers and
numerically reveal their instability.

Our presentation will be structured as follows. In section II,
we will present the model and numerical methods (the details
of which are relegated to a technical appendix). Subsequently,
in section III, we will present and discuss our numerical results. Finally,
in section IV, we will summarize our conclusions and pose questions
of interest for future studies.

\section{The model}

We consider the coupled charge/excitation-lattice model
introduced in \cite{FUENTES},
describing the
competition between linear polaronic self trapping and self-focusing effects
of a soft nonlinear potential; in dimensionless form, it can be expressed as follows:
\begin{eqnarray}
     i\dot\Psi_n = -J(\Psi_{n+1}+\Psi_{n-1}) - \chi u_n\Psi_n, \label{eq:dyn1} \\
     \ddot u_n = -V'(u_n)+\chi|\Psi_n|^2+k(u_{n+1}-2u_{n}+u_{n-1}), \label{eq:dyn1b}
\end{eqnarray}
where dots denote time derivatives, and the lattice index $n$ runs from $1$ to $N$
(the total number of lattice sites). In Eqs. (\ref{eq:dyn1})-(\ref{eq:dyn1b}),
$\Psi_n(t)$ represents the ``electronic'' degrees of freedom,
$u_n$ corresponds to the
lattice displacements (i.e., ``vibrational'' degrees of freedom), while the
parameters $J$, $k$ and $\chi$ denote, respectively, the
transfer integral, the lattice spring constant and the coupling constant between the interacting fields.
Finally, $V(u_n)$ is an anharmonic on-site potential, which, similarly to \cite{FUENTES}, is
assumed to have the form of a Morse potential:
\begin{equation}
    V(u)=\frac{1}{2}[\exp(-u)-1]^2.
\label{pot}
\end{equation}

The dynamical equations (\ref{eq:dyn1})-(\ref{eq:dyn1b}) arise from the Hamiltonian:
\begin{eqnarray}
    H &=&\sum_n\left[\frac{1}{2}\dot u_n^2+V(u_n)+\frac{k}{2}(u_n-u_{n+1})^2\right]
\nonumber
\\
&-& \sum_n \left[\chi(|\Psi_n|^2u_n)+J(\Psi_n\Psi^*_{n+1}+ {\rm c.c.})\right]
\label{hamiltonian}
\end{eqnarray}
We will seek standing-wave, as well as  genuinely time-periodic solutions
of Eqs. (\ref{eq:dyn1})-(\ref{eq:dyn1b}). To factor out
the phase invariance of Eq. (\ref{eq:dyn1}),
and clearly separate between the two classes of solutions,
we introduce the transformation:
\begin{equation}
    \Xi_n(t)=\Psi_n(t)\mathrm{e}^{-i\we t}.
\end{equation}
Then, the dynamical  equations take the following form:
\begin{eqnarray}
    i\dot\Xi_n-\we\Xi_n + J(\Xi_{n+1}+\Xi_{n-1})+\chi u_n\Xi_n = 0, \label{eq:dyn2} \\
    \ddot u_n+V'(u_n)-\chi|\Xi_n|^2-k(u_{n+1}-2u_n+u_{n-1})=0. \label{eq:dyn2b}
\end{eqnarray}
Stationary solutions now correspond to ones where all time
derivatives in Eqs. (\ref{eq:dyn2})-(\ref{eq:dyn2b}) are set to
zero. On the other hand, time periodic solutions are characterized
by a frequency of oscillation $\wb$ (or period $\Tb=2\pi/\wb$) in
the time dependence of both the electronic wavefunction $\Xi_n$ and
of the lattice displacements $u_n$. Our aim is to study the
existence and stability of such stationary and breathing polaron
solutions; the latter, adopting the terminology of  \cite{KA98,A97},
will be henceforth called ``polarobreathers''. In the results to be
presented below, we have fixed the values of the transfer integral
and lattice spring constant, namely $J=0.005$ and $k=0.13$, and let
the coupling constant $\chi$ and the polarobreather frequency $\wb$
vary. Nevertheless, in our study, we have also considered other
parameter values (such as, e.g., $k=0.065$, $k=0.26$, $J=0.0025$ and
$J=0.01$) and we have obtained qualitatively similar results to
those reported below.

In order to find either stationary or time-periodic solutions of the
dynamical equations (\ref{eq:dyn2})-(\ref{eq:dyn2b}), we use methods
based on the anti-continuous limit \cite{MA94,MA96}. In particular,
we first obtain a nontrivial steady state $u_0$ -for stationary
solutions- or an orbit of frequency $\wb$ for an isolated oscillator
$u_0(t)$-for time periodic solutions. Then, the solution at the
anticontinuous limit ($k=J=\chi=0$) is $\Xi_n=u_n=0$, except at
$n=0$, which is set to $u_0$. The coupling constants are
subsequently varied through a path-following (Newton-Raphson)
method. In the following, we first vary $\chi$ up to $0.02$, then
$J$ up to $0.005$ and, finally, $k$ up to $0.13$. Once $J$ and $k$
are fixed, $\chi$ can be freely varied. The implementation of the
anticontinuous limit has been performed in real space (using a
shooting method) and in Fourier space. Note that the first method is
less accurate than the second one, whereas the latter is rather time
consuming, particularly in the case of long chains. We have made use
of both methods in the calculations presented herein.

Once the numerically exact [up to the prescribed accuracy
O($10^{-7}$)] solutions are obtained, linear stability analysis is
performed to examine the dynamical stability of the solutions. In
the case of stationary solutions, this is implemented by imposing
normal mode perturbations (see, e.g., earlier work in
\cite{f10,MKRB03,VT00}) and obtaining the corresponding eigenvalues
of the ensuing linear matrix problem. On the other hand, in the case
of (time-periodic) polarobreather solutions, we impose
time-dependent perturbations to both the electronic and vibrational
fields and solve the resulting (linear) differential equations for
the perturbations from $t=0$ to $t=T_b$. Then, obtaining the Floquet
(or monodromy) matrix, which relates the perturbation vector at
$t=0$ to that at $t=T_b$, we compute the Floquet multipliers of the
periodic solution. The Floquet multipliers imply stability
(instability), if they do (do not) appear only on the unit circle.
The details of the numerical methods used in this work are discussed
in the Appendix.

\section{Numerical results}

In this section we illustrate the numerical results obtained for the
polarobreathers. Recall that in all calculations we have fixed
$J=0.005$ and $k=0.13$, $\chi$ being a free parameter.

\begin{figure}[tbp]
    \includegraphics[width=4.25cm]{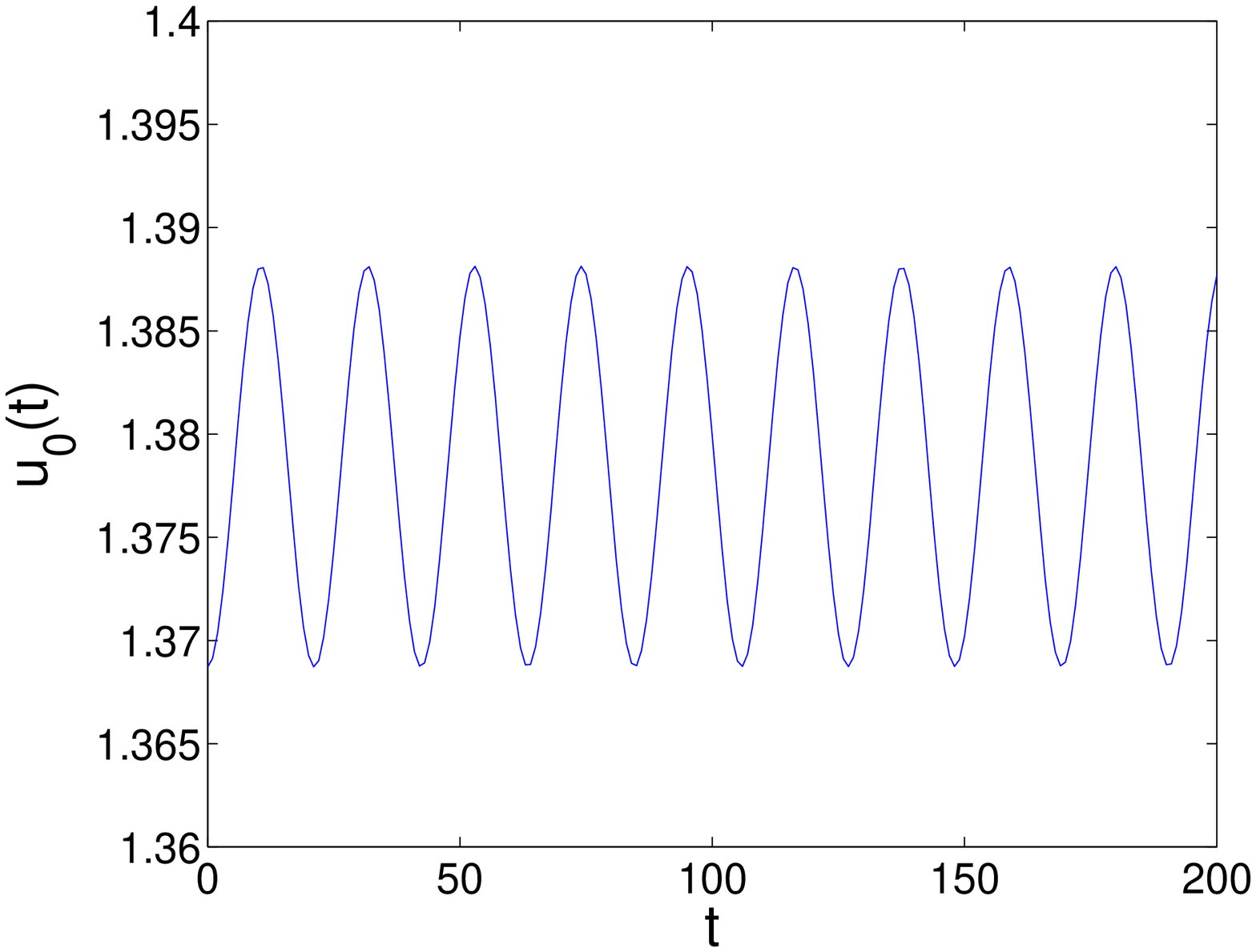}
    \includegraphics[width=4.15cm]{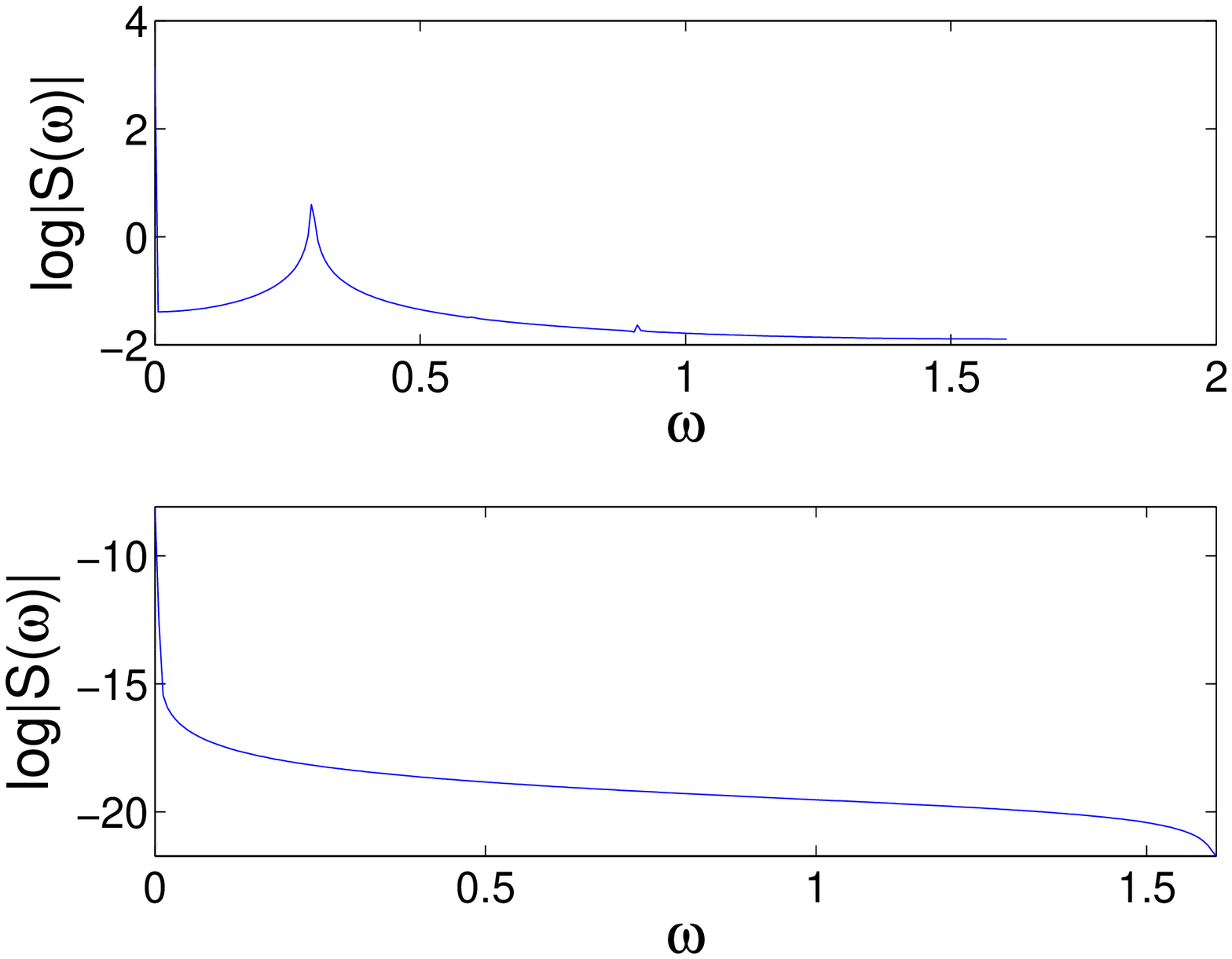}
\caption{Left panel: Evolution of the displacement of the central
particle of a slightly perturbed static polaron with $\chi=0.5$
Right panels: The Fourier spectrum of the corresponding time series
is shown (the top part corresponds to the lattice variable and the
bottom part to the electronic one). The existence of only one
frequency indicates the simple periodic nature of the solution. }
\label{fig:perturb}
\end{figure}

We first consider the stability of static polarons.
Such polarons, localized on a single lattice site, are stable.
Considering perturbations of these structures, namely
$u_n(0)=u_n^0(0)+\varepsilon\delta_{n,0}$ and
$Re(\Psi_n(0))=Re(\Psi_n^0(0))+\varepsilon\delta_{n,0}$
(where the zero superscript denotes an exact solution), it is found
that the polaron evolves into a vibrating state, which is a normal mode of the system.
This result is clearly illustrated in Fig. \ref{fig:perturb} (here we have used $\varepsilon=0.001$ and $\chi=0.5$);
note that in the relevant Fourier spectrum (right panels of the same figure
for the vibrational (top) and electronic (bottom) field at the central site),
there appears a single excited frequency.

\begin{figure}[tbp]
    \includegraphics[width=4.7cm]{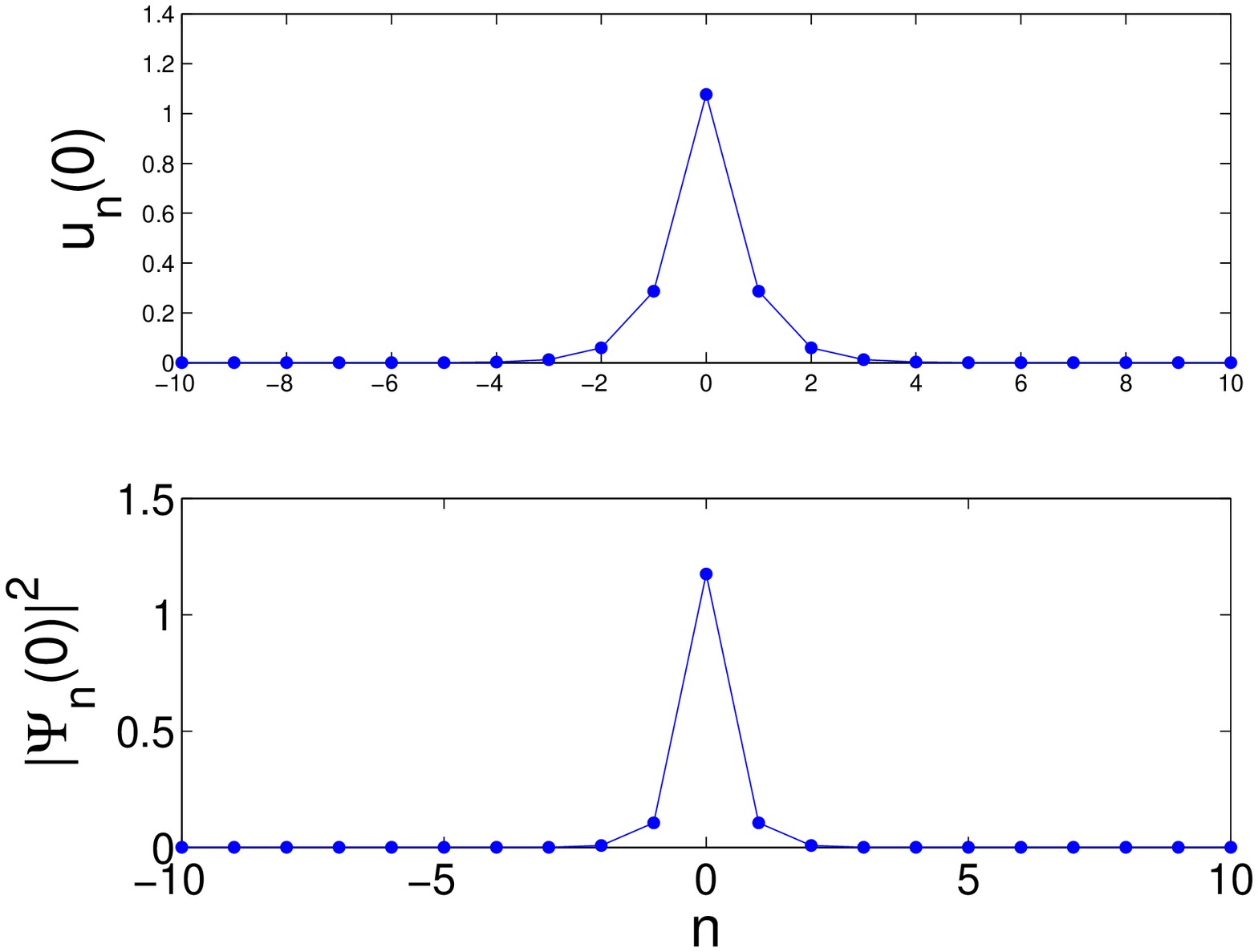}
    \includegraphics[width=3.7cm]{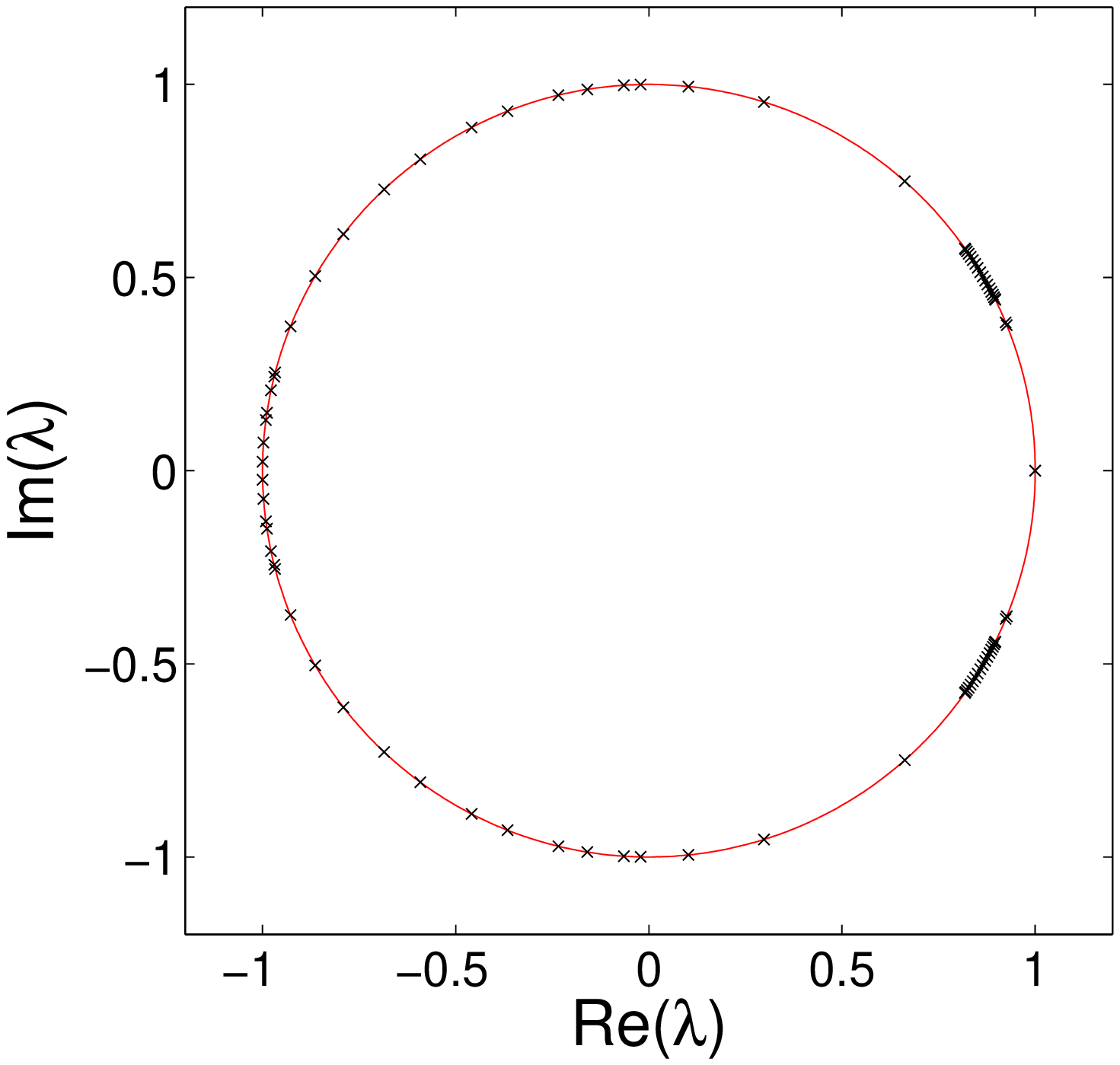}
\caption{Left panels: Profiles of the lattice ($u_n$, top) and
electronic component ($|\Psi_n|^2$, bottom) of a polarobreather with
$\wb=0.8$ and $\chi=0.15$. Right panel: Floquet spectrum of this
solution.} \label{fig:polarobreather}
\end{figure}

\begin{figure}[tbp]
\begin{center}
    \includegraphics[width=6.25cm]{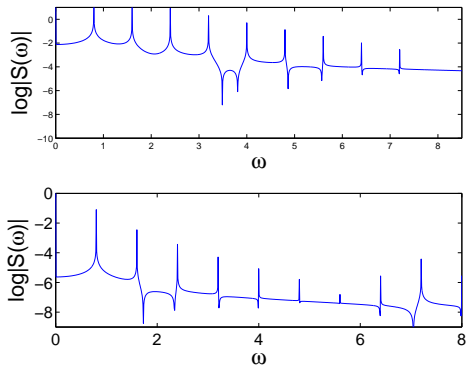}
\caption{Fourier spectrum of the polarobreather of Fig.
\ref{fig:polarobreather}. The top and bottom panels correspond to
the lattice and electronic variables respectively.}
\label{fig:Fourier}
\end{center}
\end{figure}

\begin{figure}[tbp]
    \includegraphics[width=4.2cm]{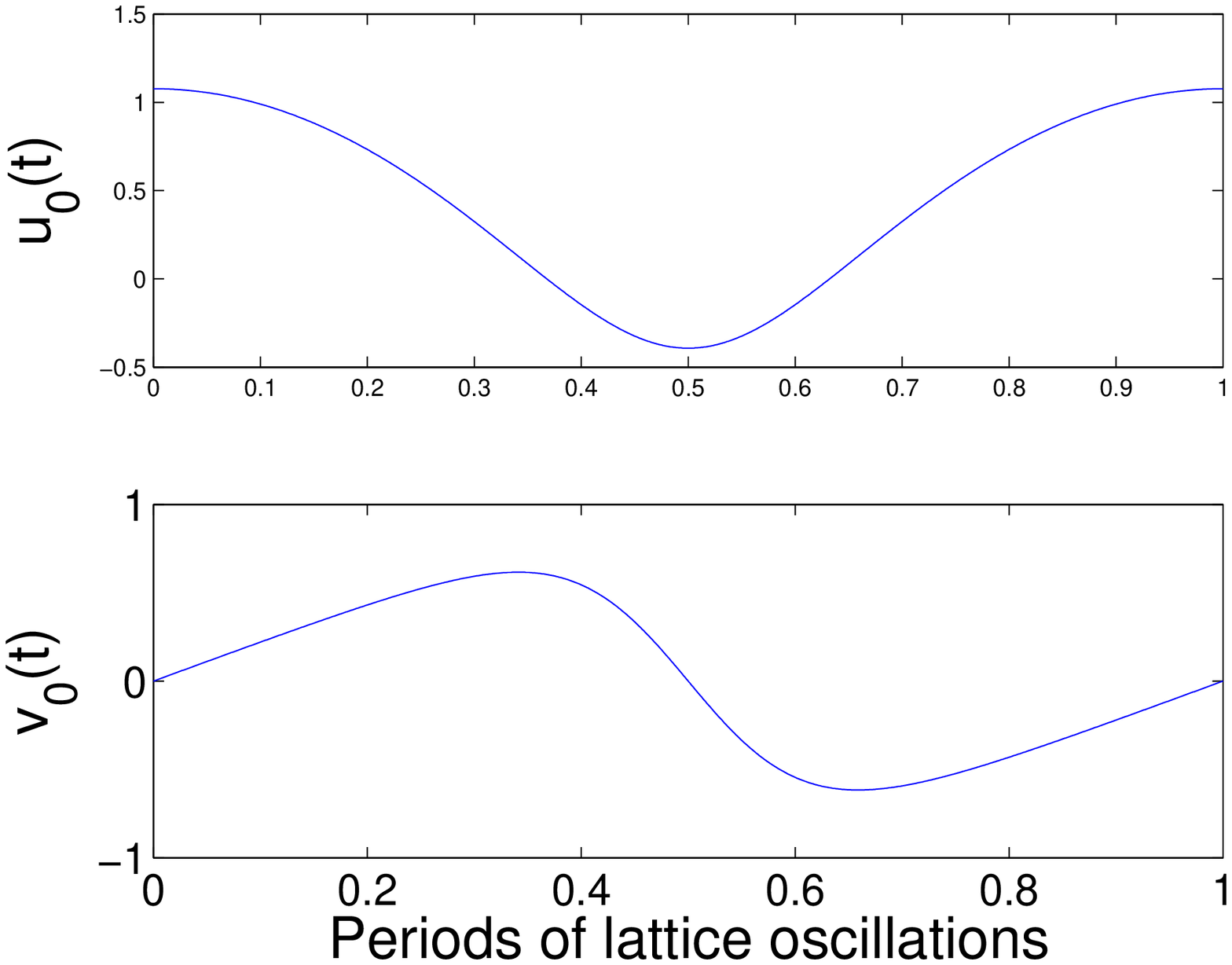}
    \includegraphics[width=4.2cm]{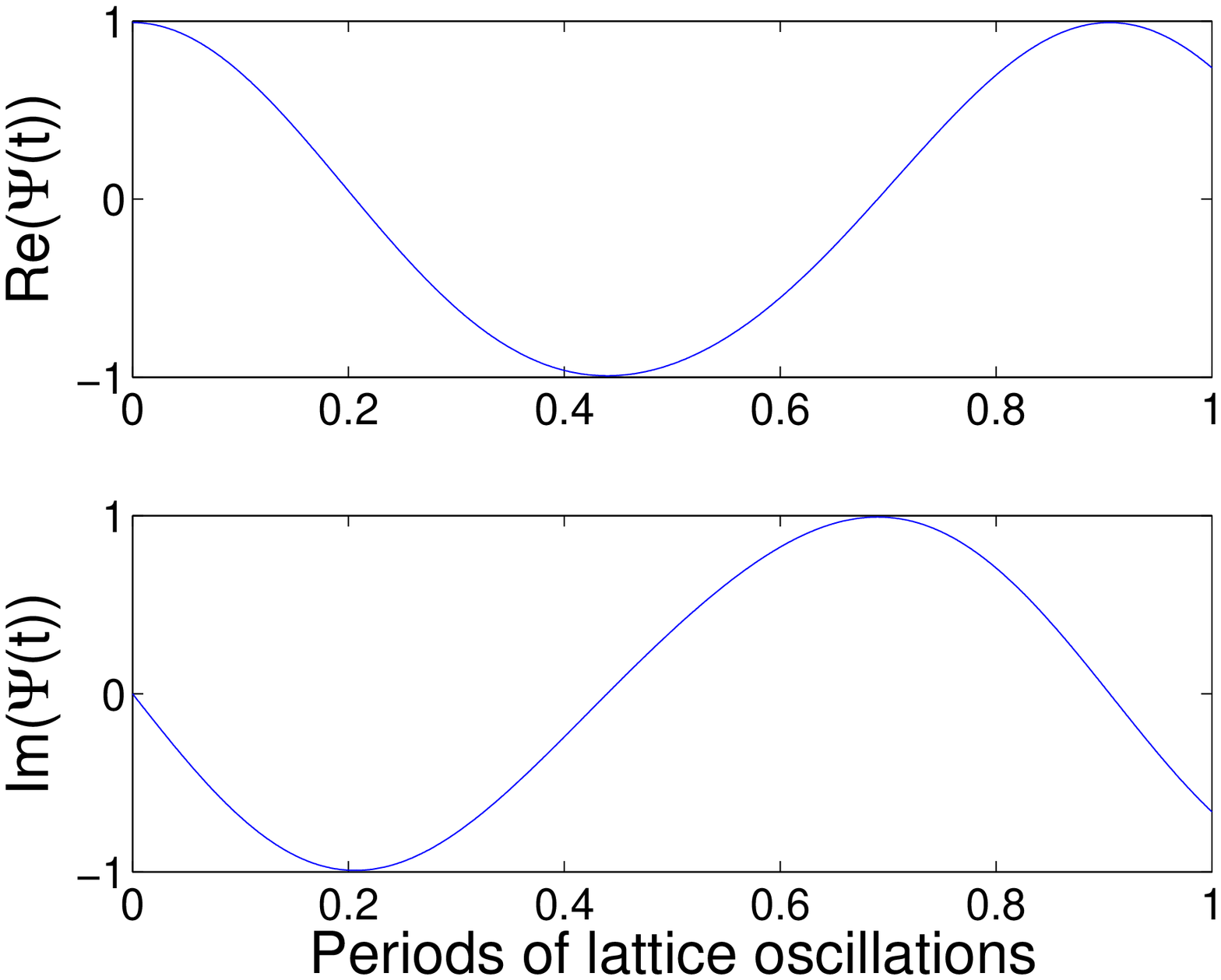}
\caption{Time evolution of the lattice (left panel) and electronic
coordinates (right panel) of a polarobreather with $\wb=0.8$ and
$\chi=0.15$. Notice that the oscillation period of the two functions
is different.} \label{fig:evol}
\end{figure}

\begin{figure}[tbp]
    \includegraphics[width=4.2cm]{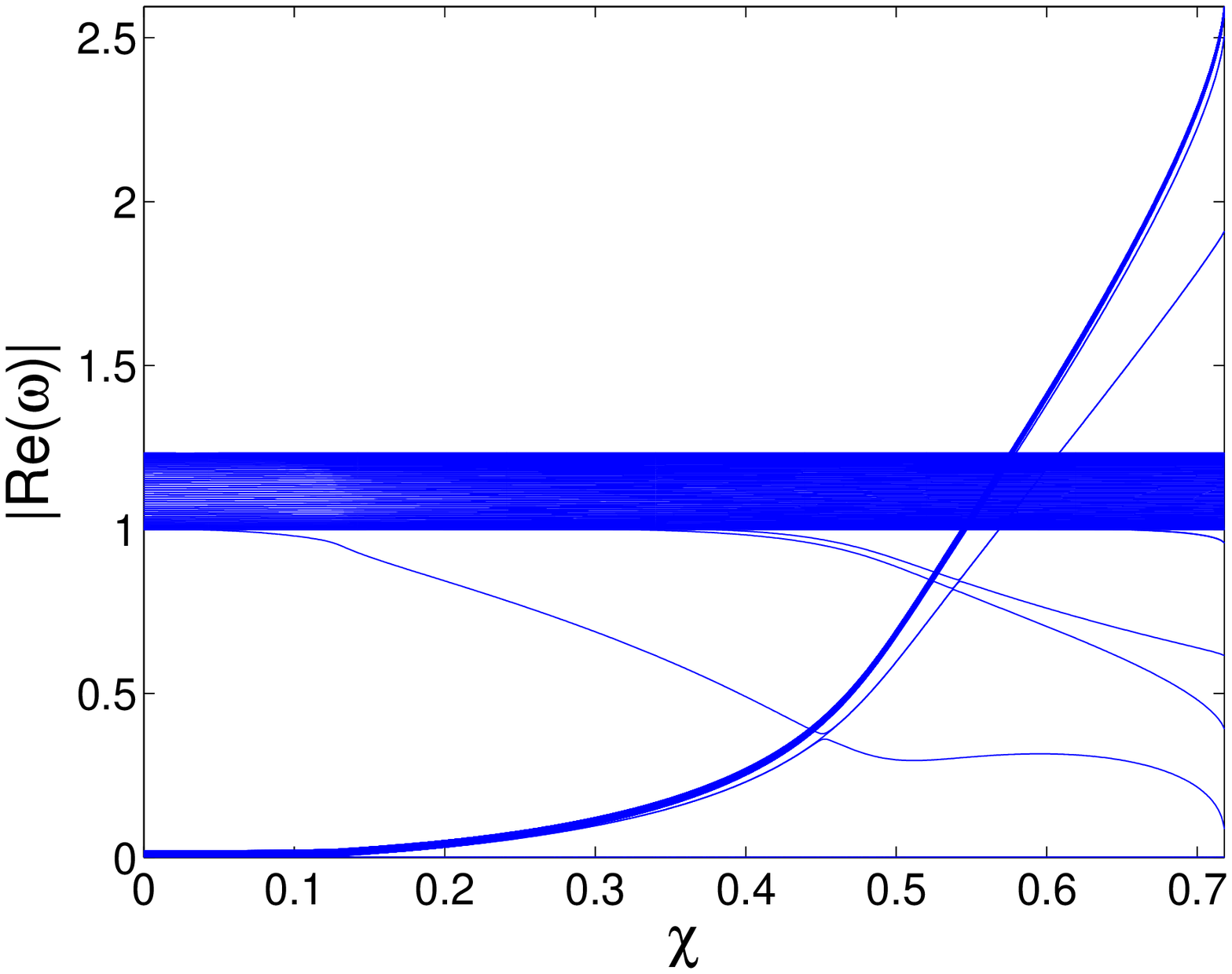}
    \includegraphics[width=4.2cm]{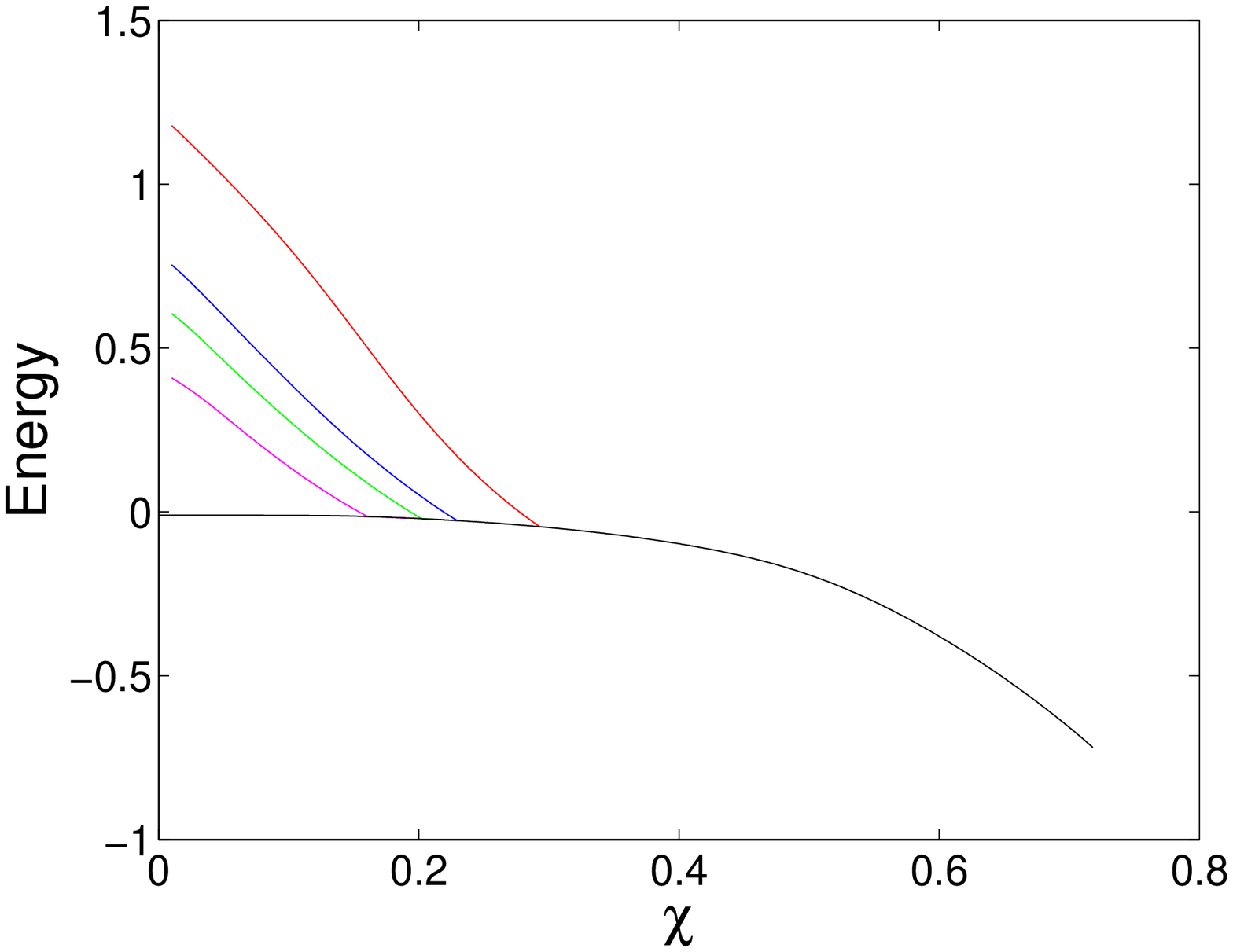}
\caption{The left panel shows the real part of the linear mode
spectrum. The right panel shows the dependence of the polarobreather
energy as a function of $\chi$ for different values of the frequency
(from left to right): $\wb=0.9, 0.84, 0.8, 0.7$. The line at the
bottom corresponds to the static polaron.} \label{fig:existence}
\end{figure}

Apart from static polarons, there exist stable, time-periodic and
spatially localized solutions within the context of the model,
namely {\em polarobreathers}. Fig. \ref{fig:polarobreather} shows an
example of these solutions along with the relevant Floquet spectrum.
Additionally, Fig. \ref{fig:Fourier} shows the Fourier spectrum of
this solution, whose nonlinear character implies the existence of
integer multiples of its frequency. The time evolution of the
lattice ($u_0(t)$) and electronic ($\Psi_0(t)$) variables of the
central particle is displayed in Fig. \ref{fig:evol}. Notice the
different period in the oscillation of the $u$ and the $\Psi$
fields, due to the additional inclusion, in the latter, of the
oscillation with frequency $\omega_e$. Polarobreathers exist as long
as the conditions of the MacKay-Aubry's theorem \cite{MA94} are
fulfilled, i.e., none of the harmonics of the polarobreather
frequency resonate with the linear modes. Based on this criterion,
an analysis of the linear modes can provide the range of existence
of polarobreathers of a given frequency. Fig. \ref{fig:existence}
shows the real part of the normal mode frequencies and indicates the
existence of a continuum band of (extended) linear modes besides
several localized modes. Since we are dealing with polarobreathers
in a soft potential, they must stem from the linear mode at the
bottom of the band of extended modes. This mode becomes localized
(i.e., bifurcates from the continuous spectrum band) at the value of
$\chi\approx0.033$ and its location varies with $\chi$. As a result,
the polarobreather frequency must be smaller than this
localized-mode frequency. This fact is indicated in the right panel
of Fig. \ref{fig:existence}. It can be the observed that the
bifurcation points correspond to the values of $\chi$ for which the
frequency of the localized mode coincides with the frequency of the
polarobreather branch.

\begin{figure}[tbp]
    \includegraphics[width=4.6cm]{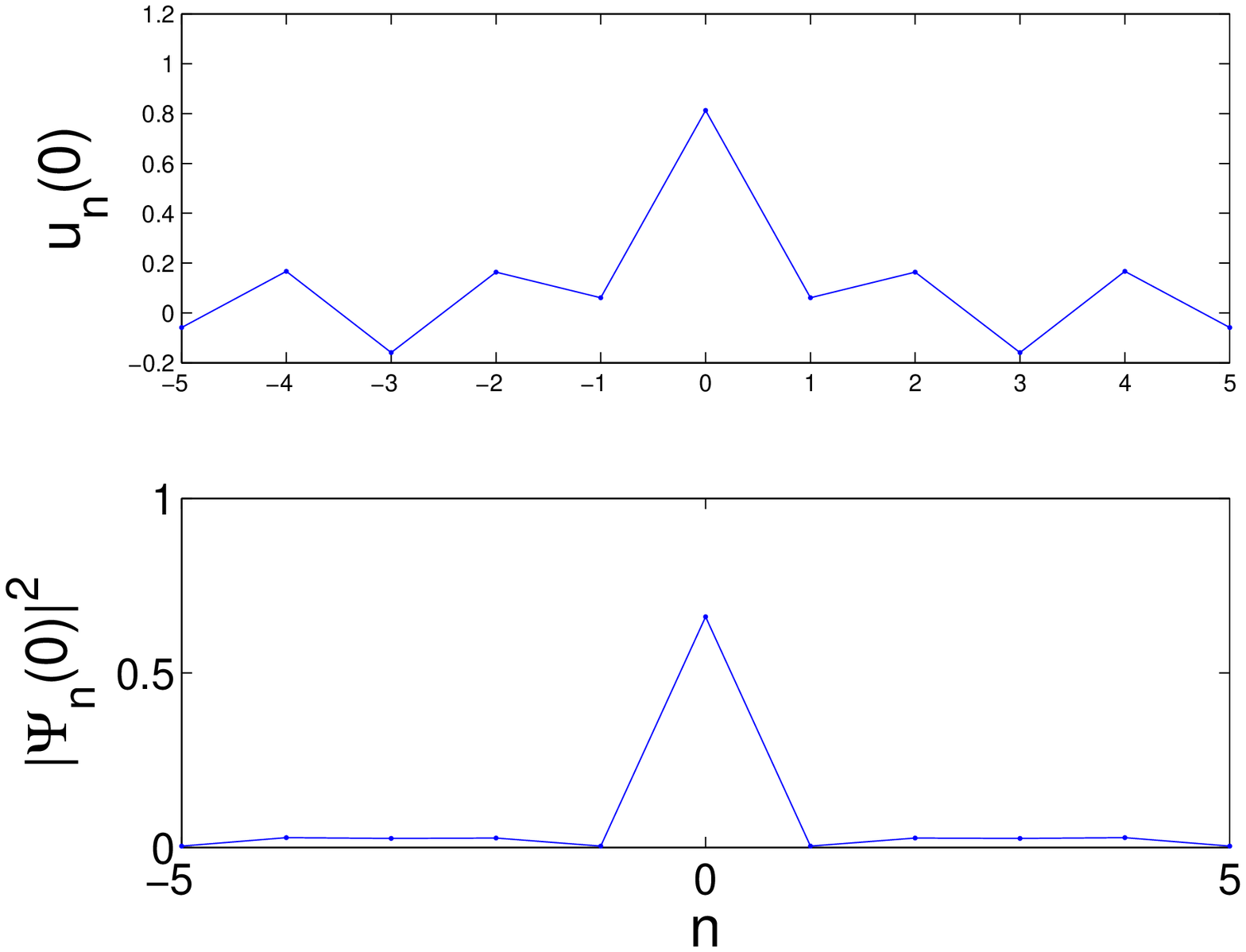}
    \includegraphics[width=3.65cm]{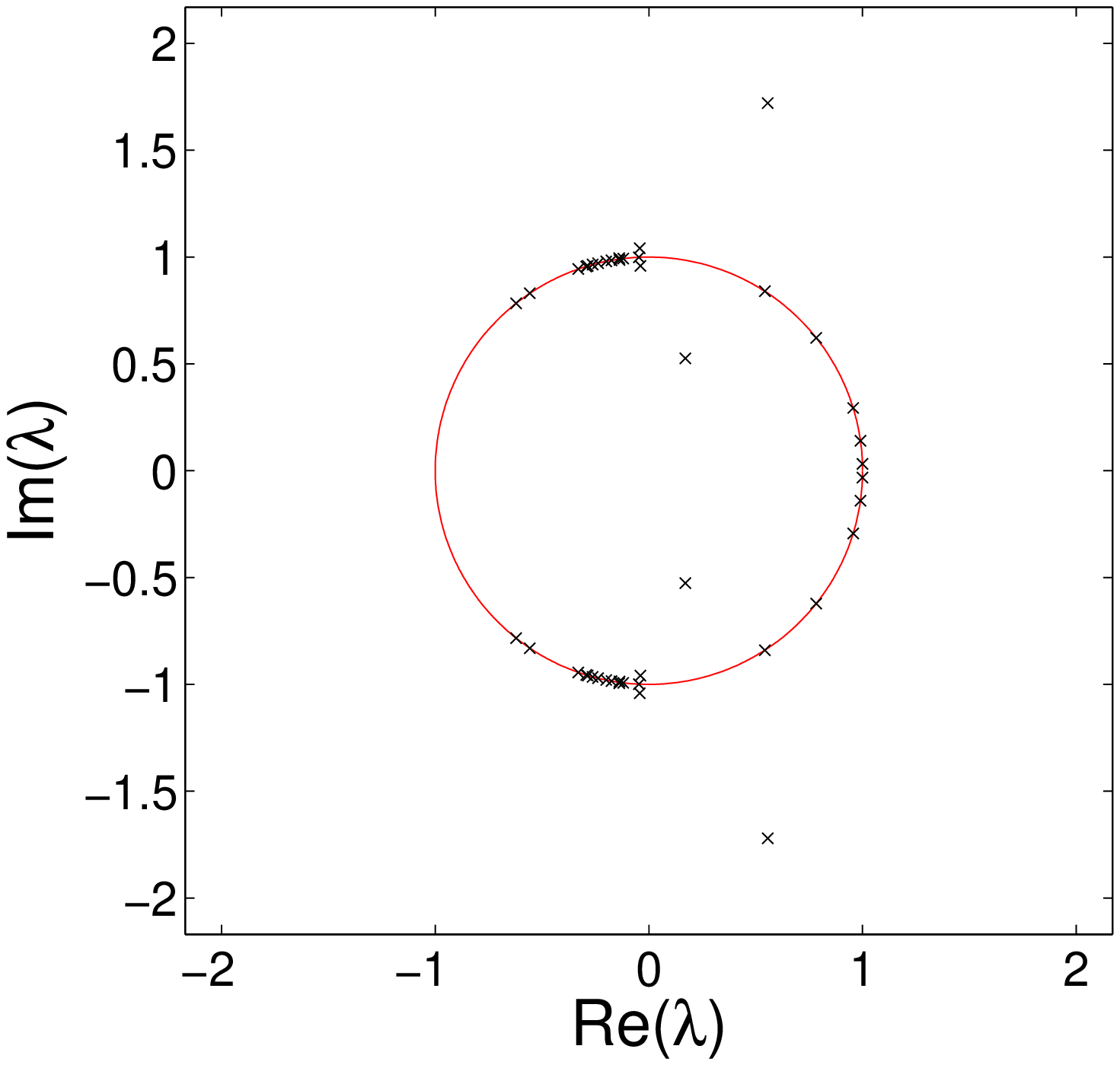}
\caption{Left panels: The profiles of the lattice (top) and the
electronic (bottom) components of a phantom polarobreather are shown
for $\wb=0.6$ and $\chi=0.15$. Right panel: Floquet spectrum of this
solution.} \label{fig:phantom}
\end{figure}

The existence of polarobreathers is also limited by the
second-harmonic resonances. In the case of finite lattices,
``phantom'' solutions can appear due to the existence of gaps in the
extended modes band \cite{phantom}. These solutions consist of a
core that vibrates with the fundamental frequency $\wb$ (or may also
be static), whereas the tails correspond to a phonon vibrating with
frequency $2\wb$. These solutions are an artifact of the finiteness
of the lattice and, moreover, they are usually unstable. An example
of a phantom polarobreather is shown in Fig. \ref{fig:phantom} (for
$\wb=0.6$ and $\chi=0.15$).

\begin{figure}[tbp]
    \includegraphics[width=4.2cm]{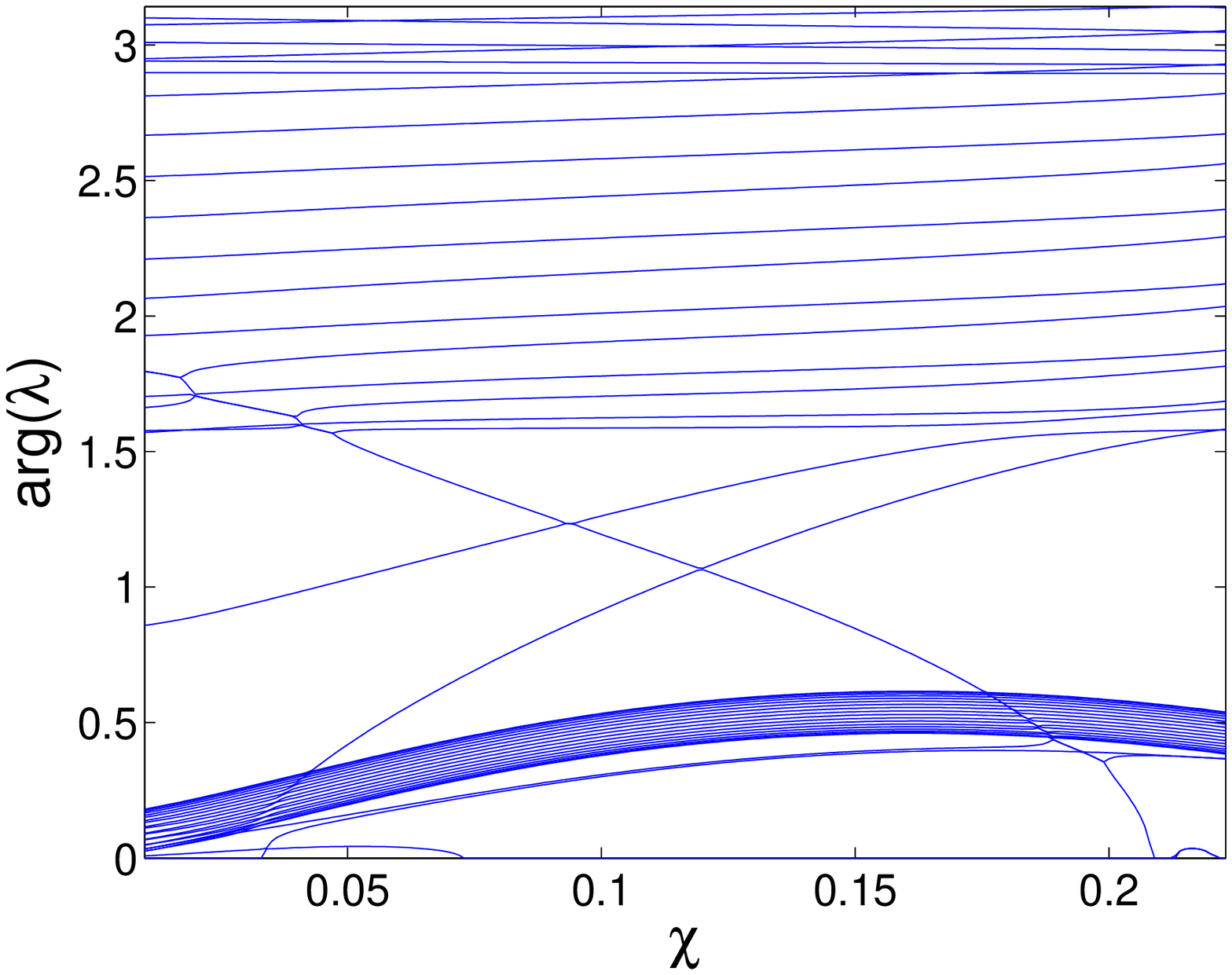}
    \includegraphics[width=4.2cm]{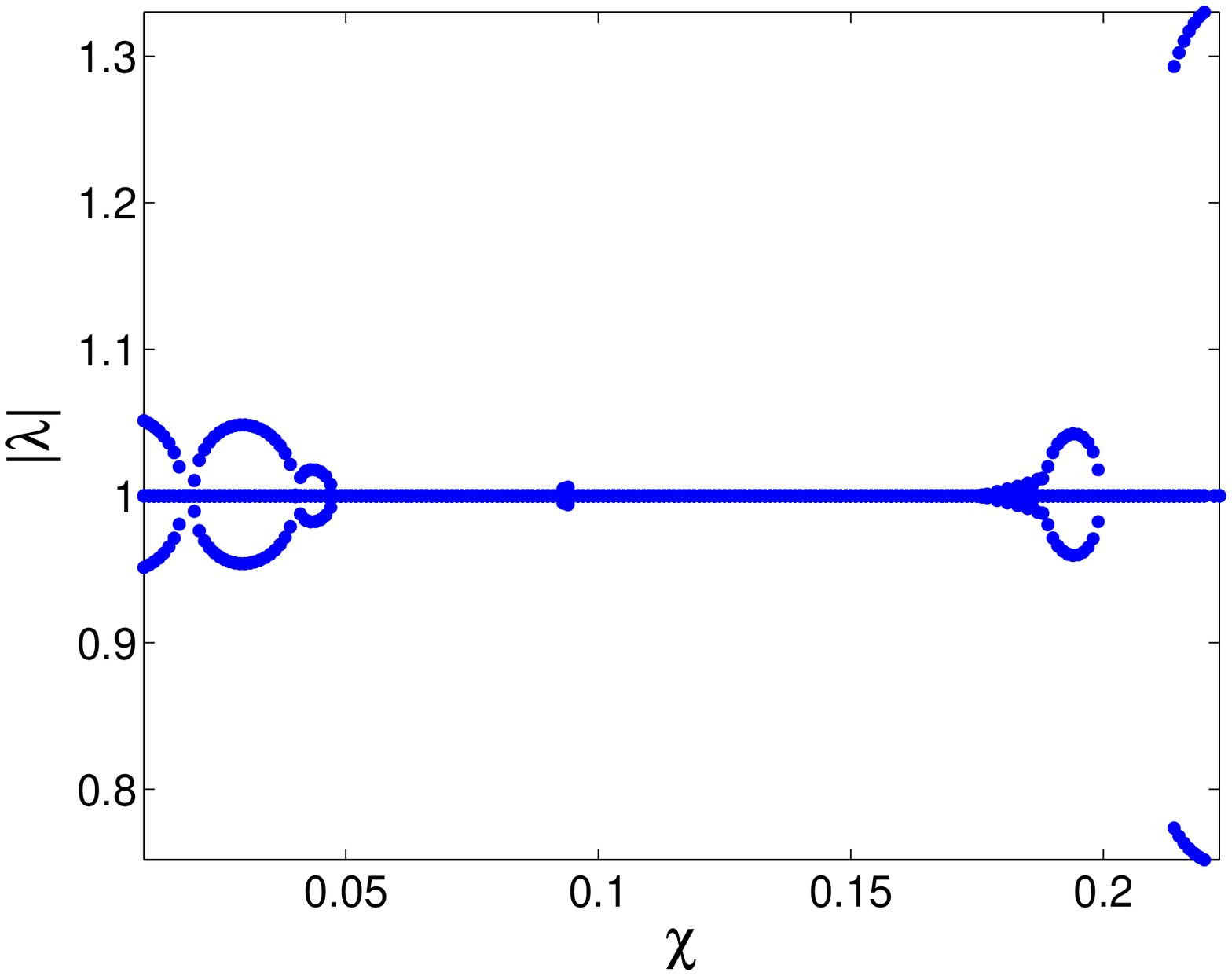}
\caption{{The argument (left) and modulus (right) of the
Floquet eigenvalues as a function of the coupling constant $\chi$
for one-site polarons.}} \label{fig:Floquet}
\end{figure}

An analysis of the stability of polarobreathers is shown in Fig.
\ref{fig:Floquet}. It is observed that, for the considered branch,
the only kind of instabilities that arise are oscillatory ones
(stemming from complex Floquet multipliers). The small growth rate
of the ensuing instabilities, however, renders the polarobreathers
relatively long-lived metastable structures when they are not
linearly stable.

\begin{figure}[tbp]
    \includegraphics[width=4.2cm]{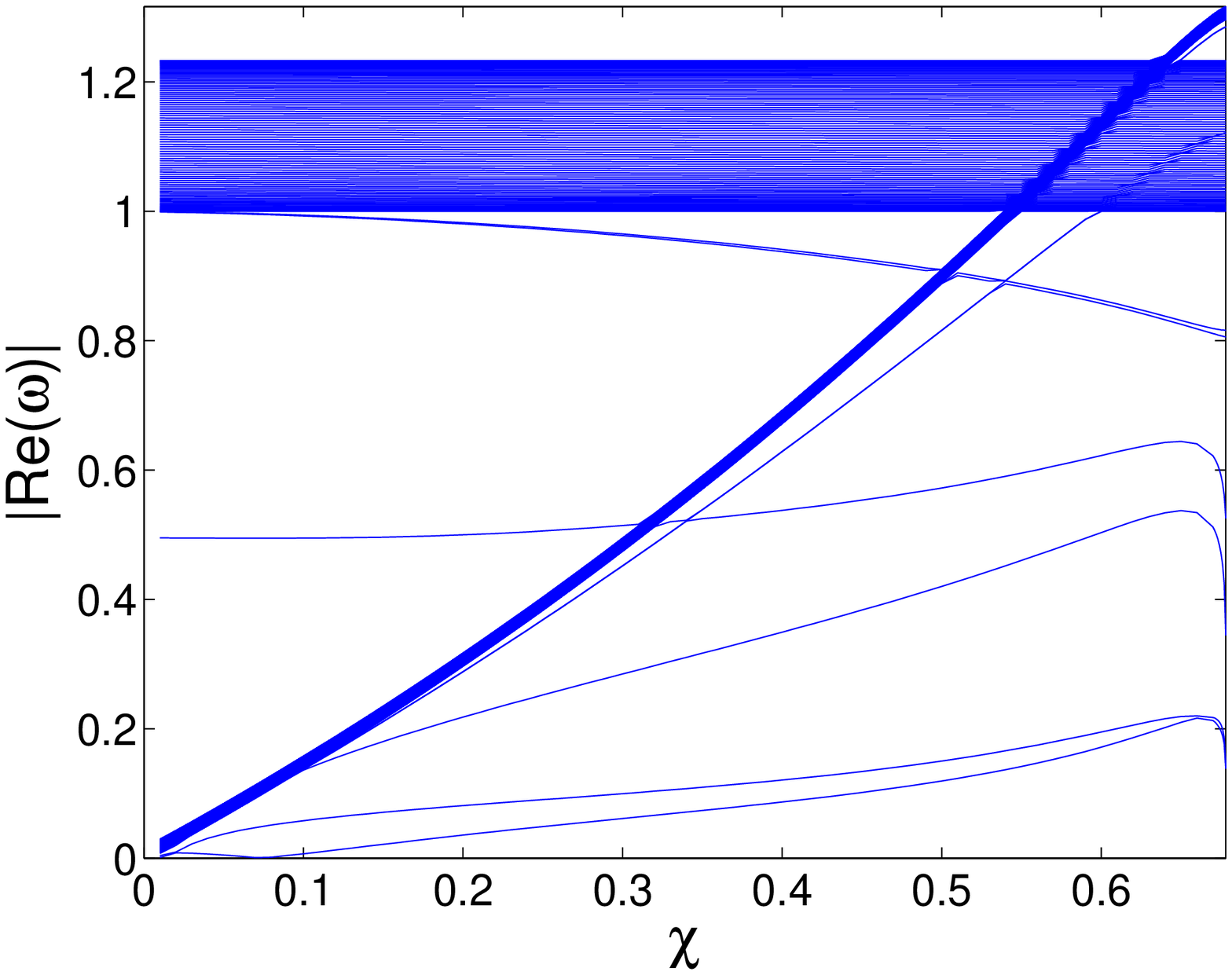}
    \includegraphics[width=4.2cm]{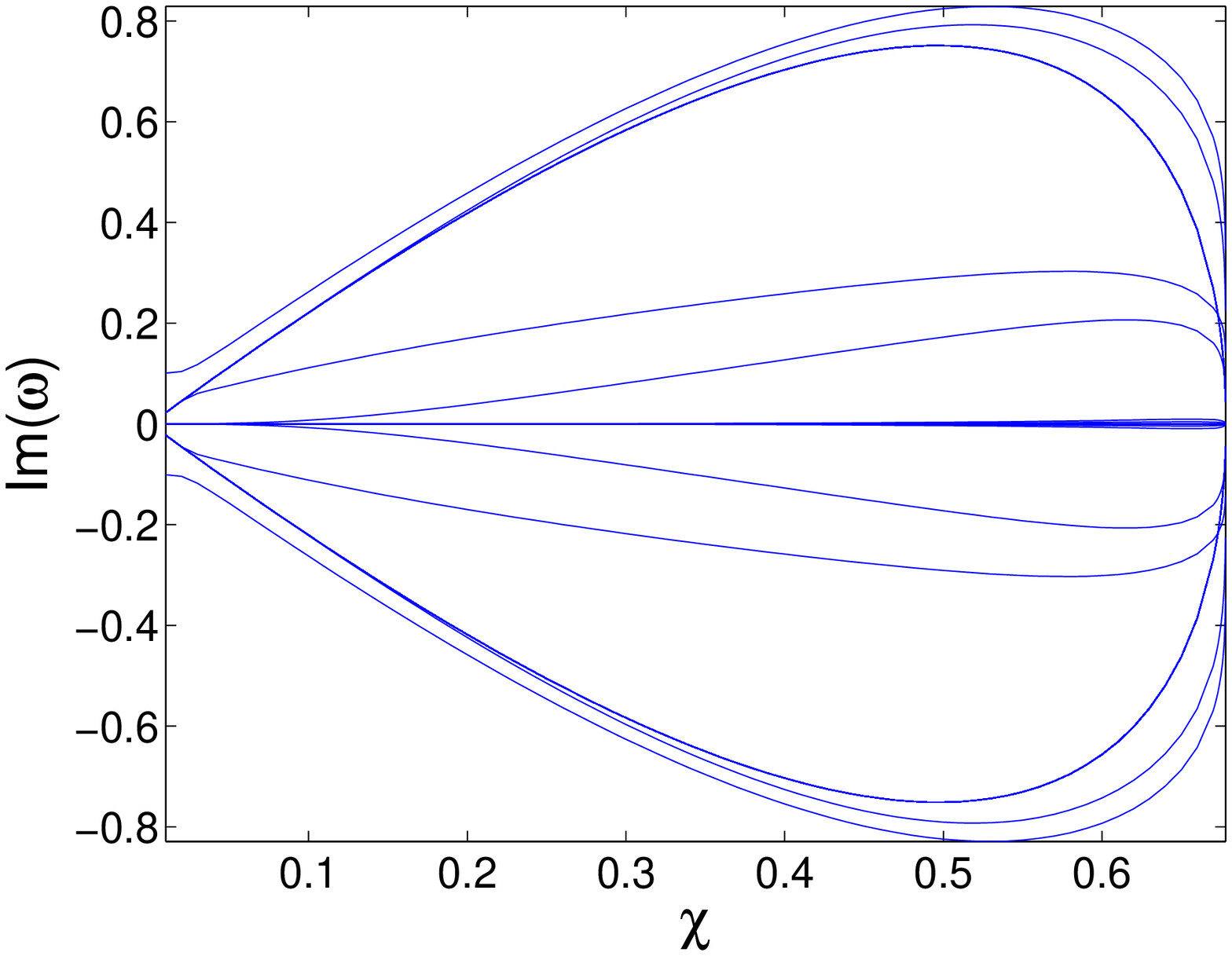}
\caption{The real (left) and imaginary (right) part of the linear
mode spectrum as a function of the coupling constant $\chi$ for
two-site polarons. } \label{fig:linear2s}
\end{figure}

Apart from one-site (site-centered) polarons, we have also
considered two-site (bond-centered) static polarons and
polarobreathers. Whereas one-site polarons are stable, two-site ones
are unstable. An example is shown in Fig. \ref{fig:linear2s}, where
the real and imaginary parts of the linear mode spectrum for
two-site polarons are illustrated.

\begin{figure}[tbp]
    \includegraphics[width=4.2cm]{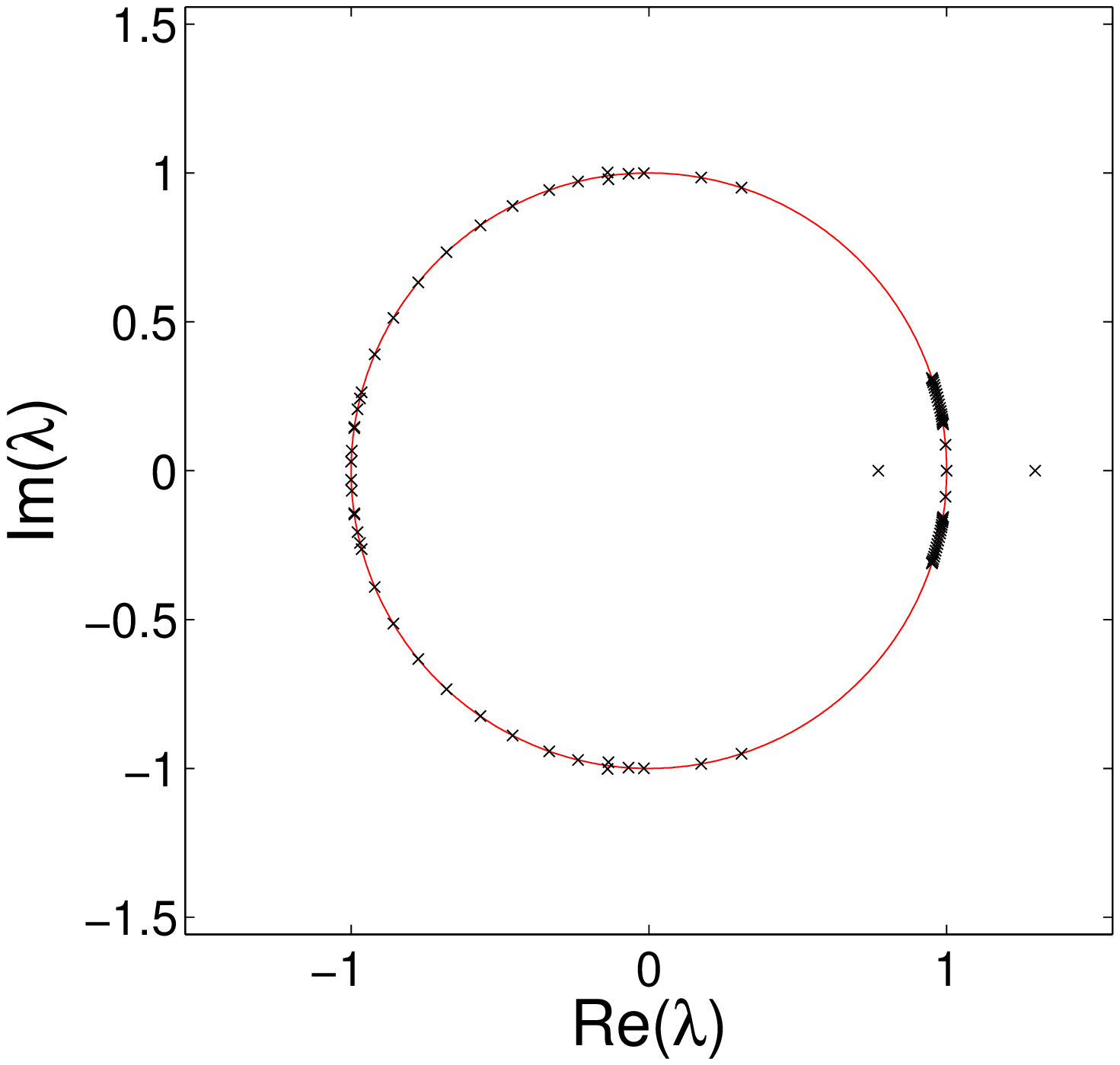}
    \includegraphics[width=4.2cm]{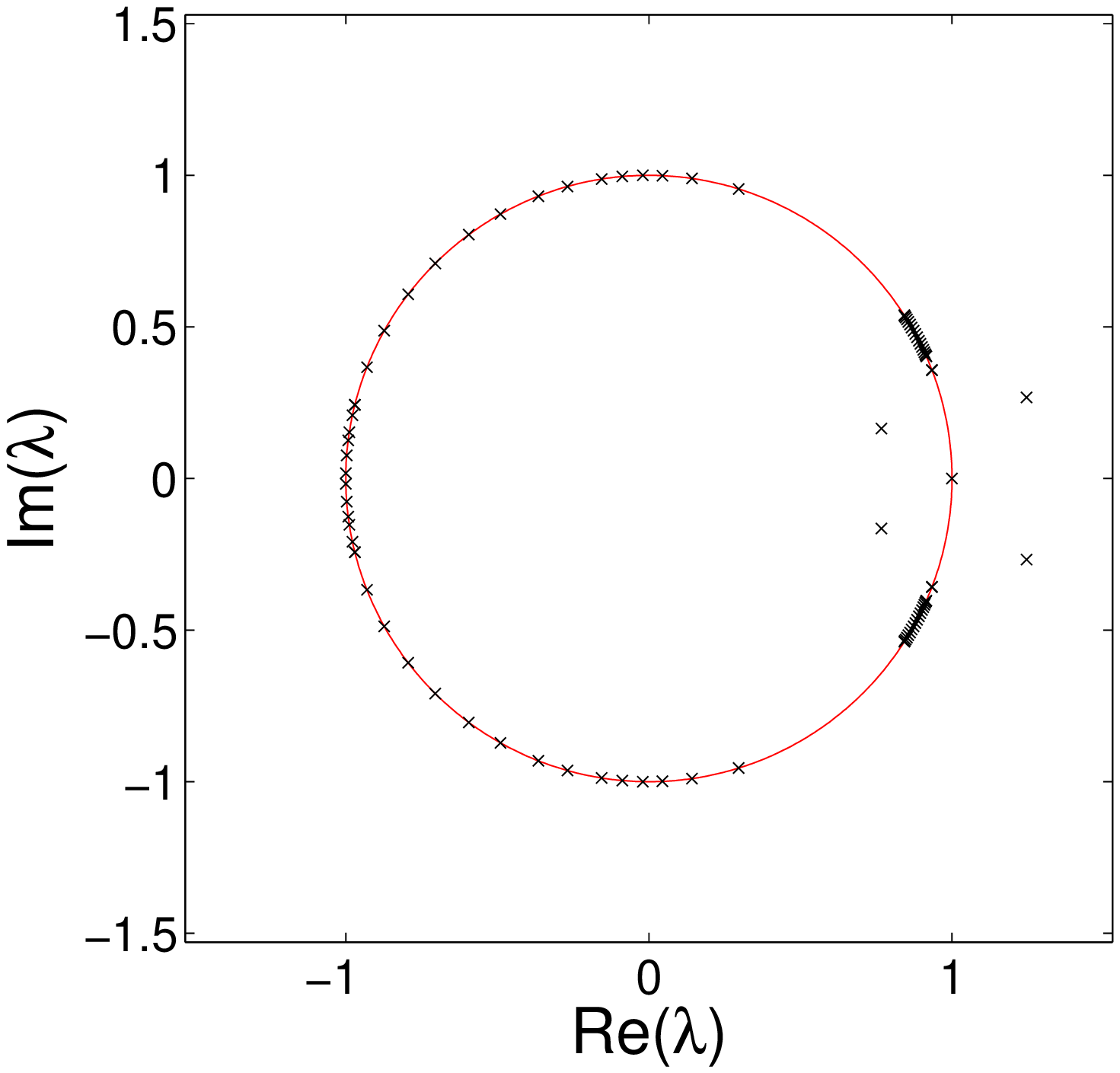}
\caption{Floquet spectra for a two-site polarobreather, subject to a
harmonic (left panel, $\chi=0.5$ ) or an oscillatory (right panel,
$\chi=0.15$) instability.} \label{fig:2spolarobreather}
\end{figure}

Similarly to their static counterparts, two-site polarobreathers are unstable.
This instability can be either exponential (i.e., the Floquet eigenvalues
responsible for the instability have zero phase) for small $\chi$, or
oscillatory (i.e., the Floquet eigenvalues responsible for the
instability have phases different from zero or $\pi$). In Fig.
\ref{fig:2spolarobreather},
examples of the Floquet spectra pertaining to a regular exponential
(for $\chi=0.5$)
and an oscillatory instability (for $\chi=0.15$) are shown.

\begin{figure}
    \includegraphics[width=6.25cm]{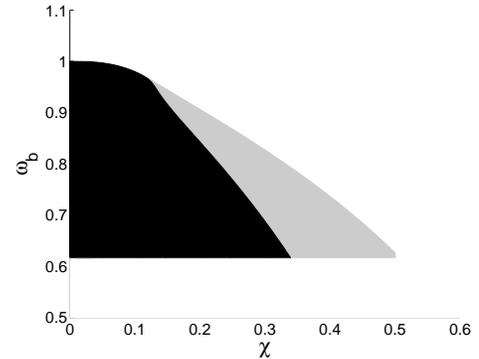}
\caption{Domain of existence of one-site (black region) and two-site
(black {\it and} gray region) polarobreathers in the $\chi$-$\wb$
parameter plane.} \label{fig:range}
\end{figure}

Two-site polarobreathers can be continued as a function of $\chi$
for $J$ and $k$ fixed. As in the one-site case, the branches for a
fixed frequency merge with a static solution branch. The domains of
existence of the one-site and two-site polarobreathers for $k=0.13$
and $J=0.005$ are shown in Fig. \ref{fig:range}. It is interesting
to note that, similarly to what happens for their static
counterparts \cite{FUENTES}, one-site polarobreathers have a
narrower domain of existence (gray region in Fig. \ref{fig:range})
than their two-site counterparts (gray and black region in Fig.
\ref{fig:range}). Given that the termination of such branches occurs
upon their collision with the stationary branch of solutions, this
trait is rather natural to expect in the present setting. A wider
domain of existence for two-site breathers was also observed in a
model with competing attractive and repulsive interaction
\cite{CAGR02}.

In order to examine the dynamical evolution of the instability of
two-site polarons, we again consider the perturbation
$u_n(0)=u_n^0(0)+\varepsilon\delta_{n,0}$ and
$Re(\Psi_n(0))=Re(\Psi_n^0(0))+\varepsilon\delta_{n,0}$, with the
value $\varepsilon=0.001$ pertaining to a static polaron. It can be
observed that, as illustrated in Fig. \ref{fig:perturb2s}, after a
transient, the polaron evolves into an anti-phase vibrating state;
note that if the two central particles were perturbed, the observed
behaviour would be similar, but the transient would be longer. If
the same perturbation is applied to a two-site polarobreather, the
behavior is somewhat reminiscent of that observed for a static
polaron. {This is illustrated in Fig. \ref{fig:perturb2s} by
displaying the temporal evolution of the energy density at the two
principal sites of the solution. The energy density is defined as:

\begin{eqnarray}
    e_n &=&\frac{1}{2}\dot u_n^2+V(u_n)+
    \frac{k}{4}\left[(u_n-u_{n+1})^2+(u_n-u_{n-1})^2\right]
\nonumber
\\
&-& \chi|\Psi_n|^2u_n-\frac{J}{2}(\Psi_n(\Psi^*_{n+1}+\Psi^*_{n-1})+
{\rm c.c.})
\end{eqnarray}

\begin{figure}
    \includegraphics[width=4.2cm]{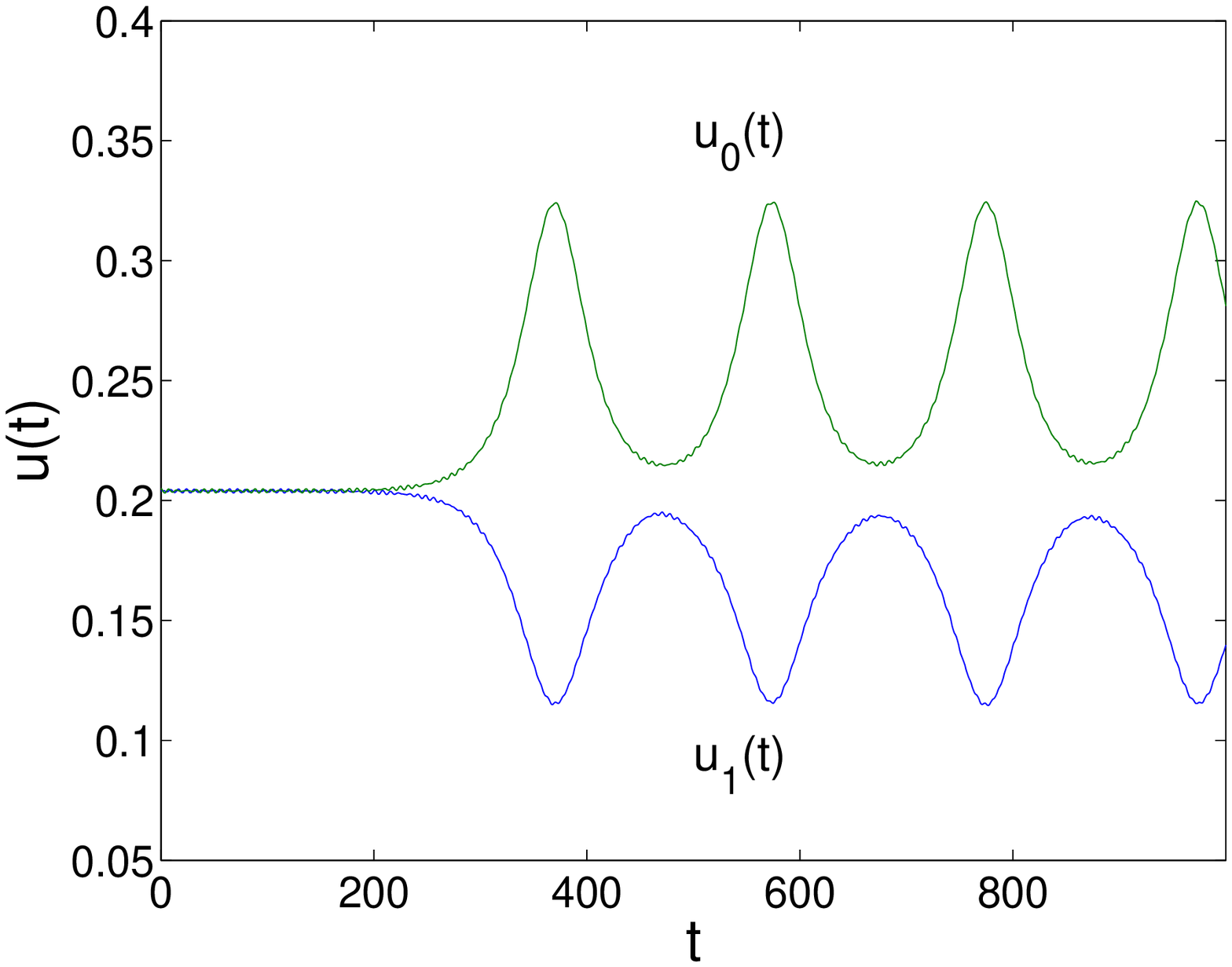}
    \includegraphics[width=4.2cm]{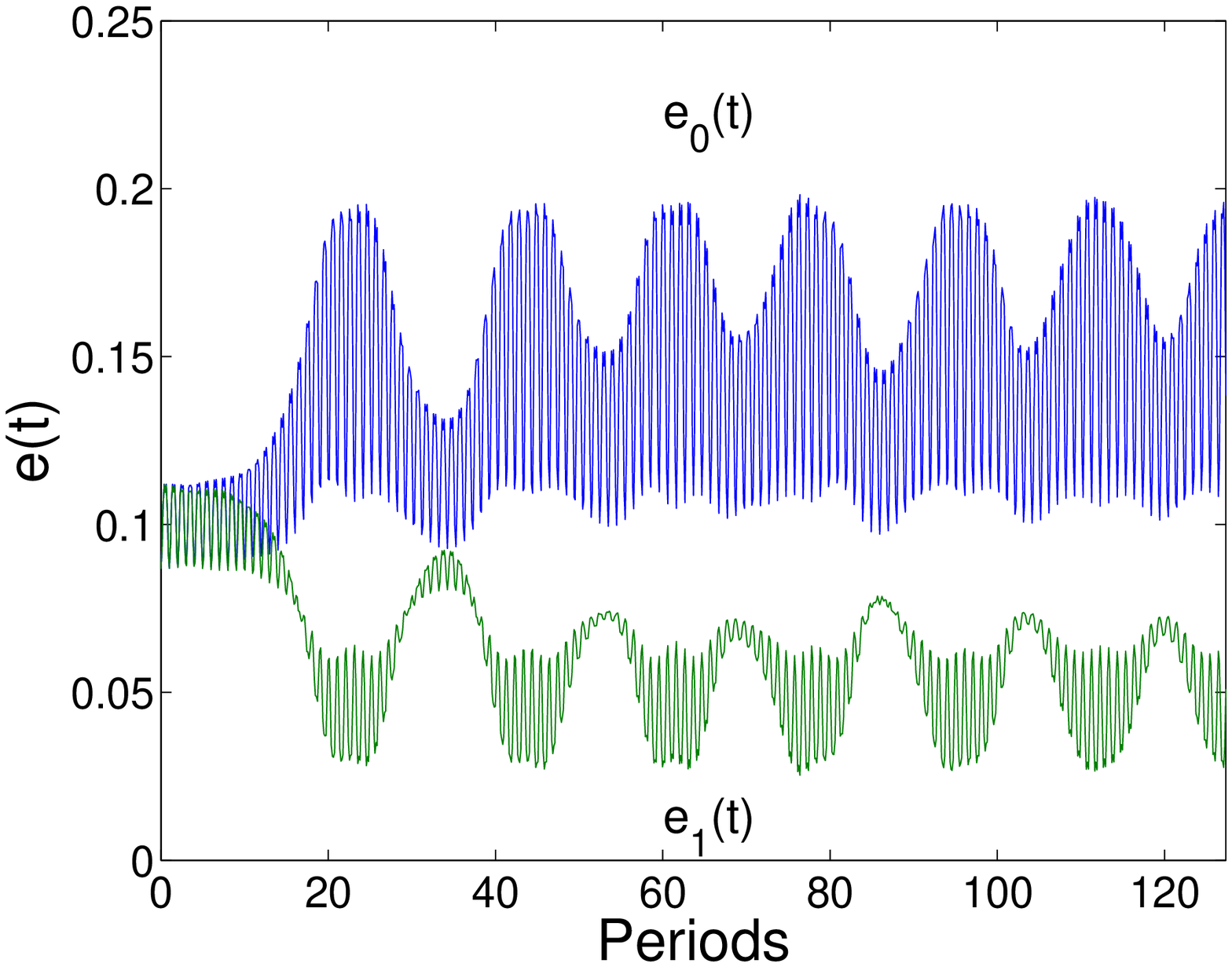}
\caption{Evolution of the central particles displacements and energy
densities, respectively, for a perturbed two-site polaron with
$\chi=0.35$ (left panel) and a two-site polarobreather with
$\chi=0.2$ (right panel).} \label{fig:perturb2s}
\end{figure}

Another interesting fact is that the static polaron branch does not
exist for $\chi>\chi_c\approx0.718$. In particular, if the solution for a polaron
close to $\chi_c$ is used as initial condition for $\chi>\chi_c)$, e.g., for
$\chi=0.72$, we have found the following type of evolution (see figure
\ref{fig:perturb2}): first, for some finite time, a continuous increase of the ``local''
energy at the central site of the solution occurs (coupled to a corresponding
decrease in the energy of the immediate neighboring sites).
Then, the system ``relaxes'' so that a quasi-breathing evolution
of the energy density is observed; however, its strong localization
at the central site of the lattice is essentially
preserved for long times.

\begin{figure}
    \includegraphics[width=4.15cm]{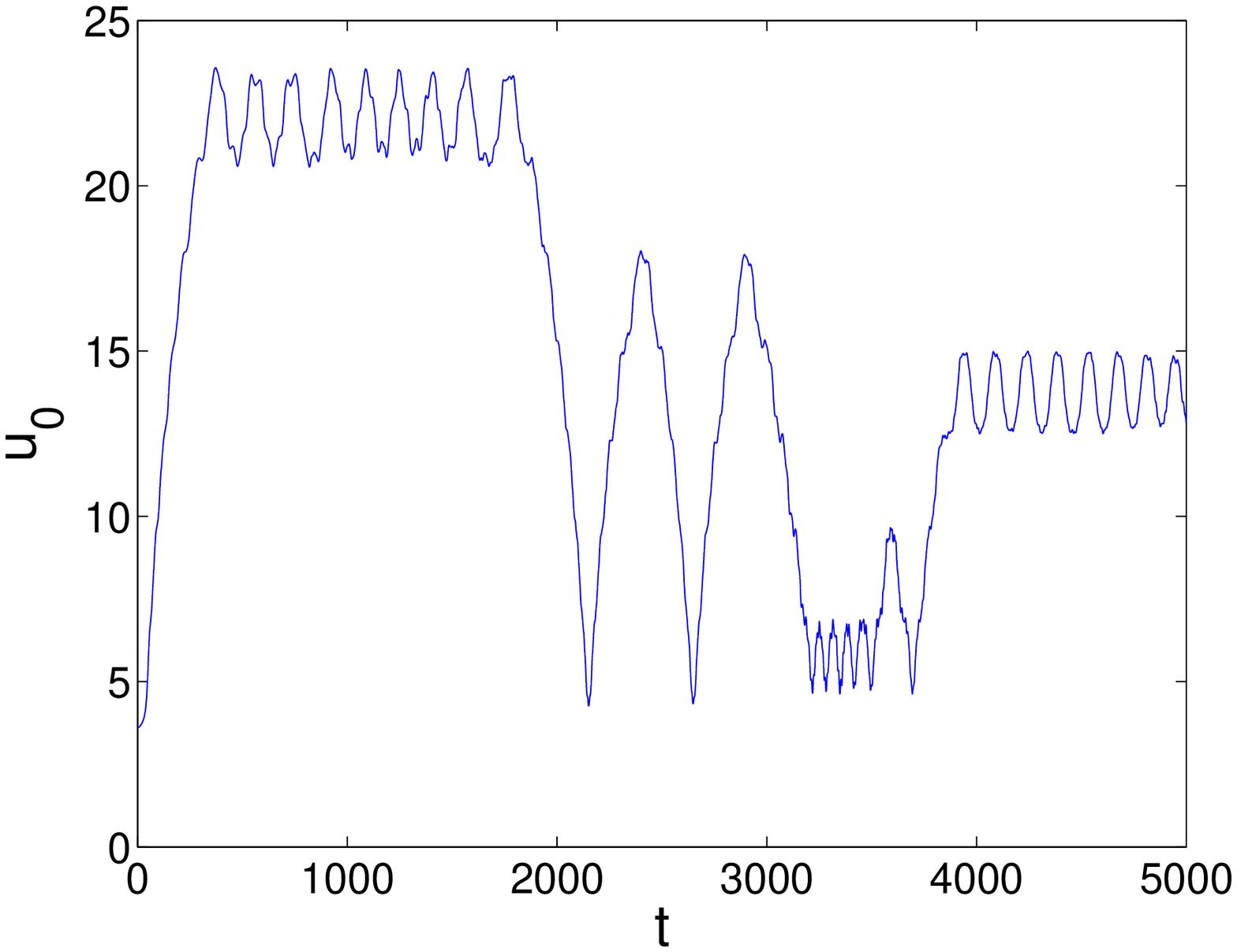}
    \includegraphics[width=4.4cm]{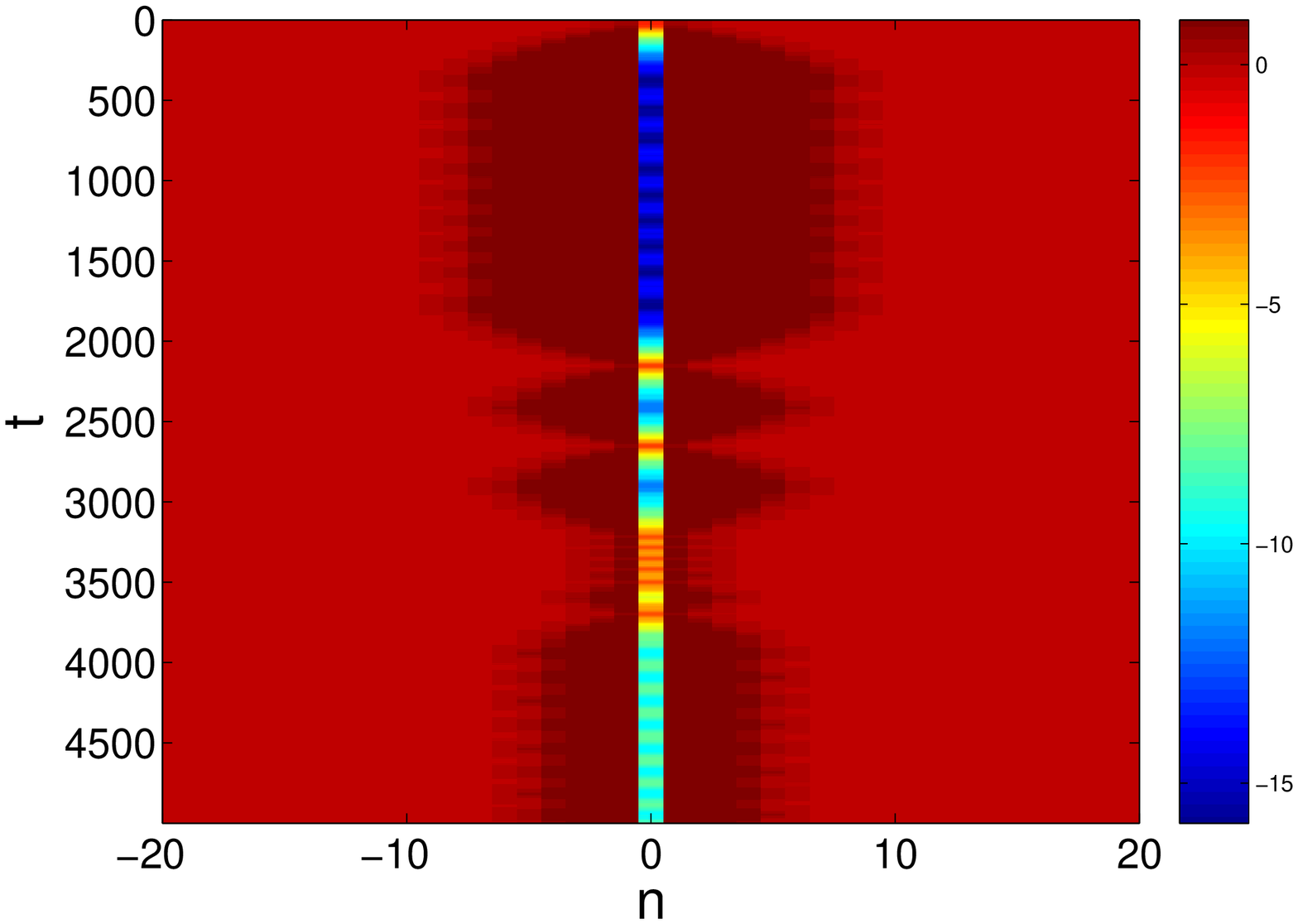}
\caption{(Color online) Left panel: Time evolution of the
displacement of the central particle at $\chi=0.72$ using as initial
condition a static polaron at $\chi=0.718$. Right panel: the energy
density plot.} \label{fig:perturb2}
\end{figure}

\section{Conclusions}

In this work, we have discussed the existence and stability
of localized solutions in a generalized Holstein model with
a soft (Morse-type) inter-particle potential. We have shown that in
addition to the standard stationary polaronic solutions
 within this model discussed elsewhere \cite{FUENTES}, there exists also
very robust breathing behavior in the form of a time-periodic type
of state: the polarobreather solutions. Using Floquet analysis, we
have illustrated the existence of parametric intervals of stability
of the latter solutions, and we have also elucidated their domain of
existence based on resonance conditions. In this respect, we have
shown that the polarobreathers cease to exist as a result of their
``collision'' with the standard polaronic branch, occurring at
frequency-dependent values of the relevant coupling parameter. We
have also found that ``phantom'' versions of such polarobreathers
exist in finite lattices but are usually unstable. We can view these
dressed polaron solutions as ``hot'' polarons (or excitons in
related exciton-vibron systems). However, the polarobreather is
dressed with a coherent phonon wavepacket (a breather), which binds
with the polaron as a composite local excitation. We have also
obtained multi-site polarons (such as the two-site polarons examined
herein) and multi-site polarobreathers (such as the two-site
polarobreathers, also studied herein), and numerically demonstrated
their instability. Finally, we have shown that the unstable
polaronic dynamics (for parameter values beyond the range of
existence of their corresponding stationary branch) may lead to
energy redistribution and eventual oscillatory dynamics for longer
timescales.

It is interesting to examine higher-dimensional generalizations of
the model and the possibility for obtaining localized solitary wave
or vortex-like solutions in such higher-dimensional discrete
contexts. Such studies are currently in progress and will be
reported elsewhere. It would also be important to assess the
significance of quantum effects in both the electronic and lattice
degrees of freedom. Initial results \cite{wang} suggest that bound
states of breathers and polarons (or excitons) may indeed persist in
appropriate regimes.

\section{Appendix: Details of Numerical Methods}

\subsection{Real space methods}

In order to analyze Eqs. (\ref{eq:dyn2})-(\ref{eq:dyn2b}) using the
real space method, it is convenient to separate the electronic
wavefunction into its real and imaginary part,
$\Xi_n(t)=\phi_n(t)+i\varphi_n(t)$, and also define the oscillation
velocity $v_n(t)=\dot u_n(t)$. In this way, we end up with a system
of $4N$ equations ($N$ is the number of particles of the system):
\begin{eqnarray}\label{eq:shoot}
    \dot\phi_n &=& -(-\we+\chi u_n)\varphi_n-J(\varphi_{n+1}+\varphi_{n-1}), \nonumber \\
    \dot\varphi_n &=& (-\we+\chi u_n)\phi_n-J(\phi_{n+1}+\phi_{n-1}), \nonumber \\
    \dot u_n &=& v_n, \nonumber \\
    \dot v_n &=& -V'(u_n)+\chi(\phi_n^2+\varphi_n^2) \nonumber \\
        &+& k(u_{n+1}-2u_n+u_{n-1}).
\end{eqnarray}
Then, if we define
$X(t)\equiv\{u_n(t),v_n(t),\phi_n(t),\varphi_n\}$, there exists a
map $\mathcal{T}$ which relates $X(0)$ and $X(\Tb)$ as
$X(\Tb)=\mathcal{T} X(0)$. The shooting (real space) method consists
in finding zeros of the map $F\equiv\mathcal{T}X(0)-X(\Tb)$
\cite{KA98} (for a detailed explanation of the shooting method, see
 \cite{MARIN}). The tangent map needed for the
application of the Newton method, $J\equiv\partial\mathcal{T}$, must
be calculated numerically.

However, in this scheme the electronic norm is not conserved and
the electronic frequency $\we$ must be known. For this reason, an
equation is added to (\ref{eq:shoot}),
\begin{equation}
    \sum_n(\phi_n^2+\varphi_n^2)-1=0,
\end{equation}
so that the norm is conserved, and a new variable $\we$ is added to
the vector $X(t)$.

\subsection{Fourier space methods}

The Fourier space methods (see details in \cite{AMM99,CAPR01}) are
based on the fact that both the electronic wavefunction and the
lattice displacements are periodic with period $\Tb$. Thus, they can
be expressed in terms of a truncated Fourier series expansion:

\begin{eqnarray}\label{eq:series}
    \Xi_n(t)&=&\sum_{k=-k_m}^{k_m} a_n^k\exp(i k \wb t), \nonumber \\
    u_n(t)&=&\sum_{k=-k_m}^{k_m} b_n^k\exp(i k \wb t).
\end{eqnarray}

Then, the dynamical equations (\ref{eq:dyn2})-(\ref{eq:dyn2b}) are reduced to a set of
$(2N+1)\times(2k_m+1)$ algebraic equations (we have also included the norm conservation),
where the variables are $Z\equiv\{a_n^k,b_n^k,\we\}$:
\begin{eqnarray}\label{eq:Fourier}
    F^k_{E,n} & \equiv & -\wb k a_n^k-\we a_n^k+\F^k(V'(u_n))+ \chi\F^k(u_n\Xi_n) \nonumber \\
              &+& J(a_{n-1}^k+b_{n+1}^k)=0, \nonumber \\
    F^k_{B,n} & \equiv & -\wb^2k^2b_n^k+\F^k(V'(u_n))- \chi\F^k(|\Xi_n|^2) \nonumber \\
              &-& k(b_{n-1}^k-2b_n^k +b_{n+1}^k)=0 \nonumber \\
    F_{N,n}   & \equiv & (\sum_n\sum_k\sum_{k'}a_n^ka_n^{k'})-1=0.
\end{eqnarray}
In the above equations, $\F^k$ denotes the Discrete Fourier Transform:
\begin{equation}
    \F^k(u)=\sum_{j=-k_m}^{k_m}u(t_j)\exp(i k \wb t_j),
\end{equation}
where $t_j$ is a sample of times that must be equally spaced according to
\begin{equation}
    t_j=\frac{2\pi j}{\wb(2k_m+1)}, \qquad j=-k_m,\ldots,+k_m,
\end{equation}
and $u(t_j)$ is calculated from the Fourier coefficients $b^k$ by
means of the Inverse Discrete Fourier Transform:
\begin{equation}
    u(t_j)=\sum_{k=-k_m}^{k_m}b^k\exp(i k \wb t_j).
\end{equation}
The Newton operator or Jacobian $J\equiv\partial F$ (with
$F\equiv\{F_{P,n}^k;F_{B,n}^k;F_N\}$ can be calculated analytically.
One of the main disadvantages of this method is that the Jacobian is
singular. Polarobreathers must be calculated using singular value
decomposition \cite{CA97}. However, this method allows the
calculation of non-time-reversible solutions.

\subsection{Normal modes and linear stability analysis}

Here, we present the normal modes and stability equations, which
were used in the results presented in the main text.

\subsubsection{Stationary Solution Stability: Normal modes}

In order to calculate the normal modes we first need to calculate a
stationary polaron from Eqs. (\ref{eq:dyn1})-(\ref{eq:dyn1b}) from the following
conditions \cite{FUENTES}:
\begin{eqnarray}
    & & \Psi_n(t)=\psi_n\exp(-i\we t), \nonumber \\
    & & \ddot u_n(t)=\dot u_n(t)=0.
\end{eqnarray}
Then, the polaron is determined by the electronic wavefunction
$\psi_n$ and the lattice displacements $y_n$.

The normal modes are introduced as perturbations to a stationary
polaron \cite{f10,MKRB03,VT00}:
\begin{eqnarray}
    & & \Psi_n(t)=[\psi_n+\epsilon_n(t)]\exp(-i\we t), \nonumber \\
    & & u_n(t)=y_n+\xi_n(t), \nonumber \\
    & & \dot u_n(t)=\dot y_n+\pi_n(t)=\pi_n(t),
\end{eqnarray}
and the dynamical equations for the normal modes are:
\begin{eqnarray}
    i\dot\epsilon_n&=&-\we\epsilon_n-\chi(y_n\epsilon_n+\psi_n\xi_n)-J(\epsilon_{n+1}+\epsilon_{n-1}), \nonumber \\
    \dot\xi_n&=&\pi_n, \nonumber \\
    \dot\pi_n&=&-V''(y_n)\xi_n+\chi\psi_n(\epsilon_n+\epsilon_n^*), \nonumber \\
    &+&k(\xi_{n+1}-2\xi_n+\xi_{n-1}).
\end{eqnarray}
Normal mode frequencies can be calculated using the following relations:
\begin{eqnarray}
    \epsilon_n(t)&=&a_n\exp(i\w t)+b_n\exp(-i\w^* t) \nonumber \\
    \xi_n(t)&=&c_n\exp(i\w t)+c^*_n\exp(-i\w^* t) \nonumber \\
    \pi_n(t)&=&d_n\exp(i\w t)+d^*_n\exp(-i\w^* t).
\end{eqnarray}
Thus, the normal modes are determined by the following linear system:

\begin{eqnarray}
    \w a_n&=&\we a_n+\chi(y_na_n+\psi_nc_n)+J(a_{n+1}+a_{n-1}), \nonumber \\
    \w b^*_n&=&-\we b^*_n-\chi(y_nb^*_n+\psi_nc_n)-J(b^*_{n+1}+b^*_{n-1}), \nonumber \\
    \w c_n&=&-id_n, \nonumber \\
    \w d_n&=&i[V''(y_n)c_n-\chi\psi_n(a_n+b^*_n) \nonumber \\
    &+& k(c_{n+1}-2c_n +c_{n-1})].
\end{eqnarray}

As the normal mode operator is non-Hermitian, the frequencies are in general
complex numbers. If non-real frequencies exist in the spectrum, then the
stationary polaron is unstable.

\subsubsection{Polarobreather Stability: Floquet Spectrum}

Linear stability can be studied via a Floquet analysis. This
analysis is performed by linearizing the dynamical equations around a
polarobreather. Then, the analysis performed for the normal modes
cannot be extended to the stability, as the electronic wavefunction
does not possess gauge invariance. Thus, we introduce a perturbation
$\Omega\equiv\{\alpha_n,\beta_n,\xi_n,\pi_n\}$ to Eqs.
(\ref{eq:shoot}), which has the form:
\begin{equation}
    \phi_n=\phi_n^0+\alpha_n, \,\,
    \varphi_n=\varphi_n^0+\beta_n, \,\,
    u_n=u_n^0+\xi_n, \,\,
    \pi_n=\dot\xi_n,
\end{equation}
where the superscript zero denotes the polarobreather solution.
Then, to leading-order approximation (i.e., keeping only the linear
terms), the stability equations are:
\begin{eqnarray}
    \dot\alpha_n&=&-(-\we+\chi
    u_n^0)\beta_n-\chi\varphi_n^0\xi_n-J(\alpha_{n+1}+\alpha_{n-1}),
    \nonumber \\
    \dot\beta_n&=& (-\we+\chi
    u_n^0)\alpha_n+\chi\phi_n^0\xi_n+J(\beta_{n+1}+\beta_{n-1}),
    \nonumber \\
    \dot\xi_n&=&\pi_n, \nonumber \\
    \dot\pi_n&=&2\chi(\phi_n^0\alpha_n+\varphi_n^0\beta_n)-V''(u_n^0) \nonumber \\
    &-& k(\xi_{n+1}-2\xi_n+\xi_{n-1}).
\end{eqnarray}
The Floquet operator (or monodromy matrix) $\mathcal{M}$ relates the vector
$\Omega$ at $t=0$ and $t=\Tb$ as follows:
\begin{equation}
    \Omega(\Tb)=\mathcal{M}\Omega(0).
\end{equation}
A polarobreather is stable if all the eigenvalues of the monodromy matrix
lie on the unit circle \cite{A97}.

\noindent {\bf Acknowledgements} One of us (J.C.) acknowledges
financial support from the MECD/FEDER project FIS2004-01183.

\end{document}